\documentclass[11pt]{article}
\usepackage{amssymb}
\usepackage{amsmath}
\usepackage{geometry}
\usepackage[onehalfspacing]{setspace}

\usepackage{graphicx}
\usepackage{pdfpages}

\begin{document}

\title{Tight Guarantees in the Commons}
\author{Anna Bogomolnaia\thanks{%
University of Glasgow and CNRS, Paris} and Herv\'{e} Moulin\thanks{%
University of Glasgow}}
\date{27/05/2026}
\maketitle

\begin{abstract}
In our context-free model of a commons, the function$\mathcal{W}$ transforms
the profile of the agents' types $(x_{1},..,x_{n})$ to a freely transferable
output $\mathcal{W}(x_{1},..,x_{n})$ that they must share fairly. We expand
the ubiquitous concept of \textit{endogenous fair shares} to include both a
lower and an upper bound on agent $i$'s share at the interim stage where $i$
only knows its own type $x_{i}$. Two functions $(g^{-},g^{+})$ form a pair
of tight guarantees if 1) they satisfy the system of inequalities $%
\sum_{1}^{n}g^{-}(x_{i})\leq \mathcal{W}(x)\leq \sum_{1}^{n}g^{+}(x_{i})$
for all profiles, and 2) the interval $[g^{-}(x_{i}),g^{+}(x_{i})]$ is
inclusion minimal across all types.

For super (resp sub) modular functions 1) the \textit{Unanimity }share%
\textit{\ }$\frac{1}{n}\mathcal{W}(x_{i},x_{i},..,x_{i})$ is the unique
tight upper (resp lower) guarantee, 2) two \textit{Stand Alone} shares $%
g(x_{i})=\mathcal{W}(x_{i},\overbrace{x_{0},..,x_{0}})-\frac{n-1}{n}\mathcal{%
W}(\overbrace{x_{0},..,x_{0}})$ (where $x_{0}$ is the smallest or largest
type) bracket all tight guarantees on the other side of Unanimity, 3) serial
cost sharing implements the Unanimity and Stand Alone guarantees.

In applications to specific microeconomic models, tight guarantees vindicate
or dismiss familiar deterministic sharing rules and suggest new ones with a
clear normative interpretation. Our examples include joint production with
substitute or complementary inputs, allocating an indivisible good and cash
transfers, sharing the cost (or benefit) of the variance or the spread of
types, the waiting cost in a queue, and more.
\end{abstract}

\section{Introduction and the punchline}

The utilisation of a commons is a mechanism design problem already mentioned
by Aristotle, as private incentives result in the mis-utilisation of these
resources (\cite{Ha}). An allocation aligning incentives and efficiency (%
\cite{Se}, \cite{We}, \cite{Is}) must in addition deliver a \textit{fair}
utilisation of these resources because they are the common property of their
users. In parallel to the analysis of empirical remedies to the tragedy (%
\cite{Ost}), a substantial theoretical, and at least partly axiomatic
literature discusses competing interpretations of fairness in a variety of
commons instances: cooperative production (\cite{Ma}, \cite{FrMo}, \cite%
{FlMa}, \cite{Roe}), pricing in public enterprise (\cite{Ba}, \cite{Fa}, 
\cite{BPW}), queuing and scheduling (\cite{She},\cite{Man}, \cite{Chu}),
water management \cite{AS}), etc..

Here we assume away the efficiency and incentive compatibility issues in a
streamlined \textit{context-free} model of the commons. We propose a new
axiomatic interpretation of fairness that does not depend on the nature of
the resources.\footnote{%
Another context-free axiom is Equal Treatment of Equals property; a
context-dependent one is Envy-Freeness.}

Each agent $i$ is responsible for choosing a \textit{type} $x_{i}$ and a
profile of types $x=(x_{1},..,x_{n})$ generates a freely transferable output 
$\mathcal{W}(x)$ of which agent $i$ receives the share $y_{i}$, so that $%
\sum_{1}^{n}y_{i}=\mathcal{W}(x)$. The function $\mathcal{W}$ is the
commons, the common property resource they share. The agents may or may not
have preferences over the tradeoffs between type $x_{i}$ and output share $%
y_{i}$, but in any case they play no role in the model. Our last essential
assumption is that the set of possible types is an interval common to all
agents, and the technology $x\rightarrow \mathcal{W}(x)$ is symmetric in the 
$n$ inputs. Therefore only the differences in individual types justifies
giving them unequal shares of the output.

The model acquires economic meaning in specific examples. A familiar one is
team work and partnerships: each agent provides an input $x_{i}$ (effort) to
the collective production process and receives a share $y_{i}$ of the output 
$\mathcal{W}(x)$ (\textbf{\cite{MoSh}, \cite{RoSi}, \cite{Sp}}): Example
5.2. Or a type is the size of agent $i$'s job and $y_{i}$ the waiting time
until its completion: Example 6.1. Or $x_{i}$ is the level of a polluting
activity and $y_{i}$ a share of the cleaning costs: Example 5.1. Or $x_{i}$
is a location on a line and $y_{i}$ a share of the cost of the spread (or
variance) of the profile of locations: Examples 6.2 and 5.3.

\paragraph{Two-sided fair shares}

Ever since its introduction in the cake cutting model (\cite{St}, \cite{Ku} 
\cite{DS}), the concept of \textit{endogenous fair share }is omnipresent in
the fair division literature, as a guaranteed worst case utility independent
of the potentially adversarial types of other agents,. It is arguably its
most important idea because of its versatility and universality.

In our model it suggests two symmetric definitions. If the output of $%
\mathcal{W}$ is a benefit, a fair share is a \textit{lower guarantee}
function $x_{i}\rightarrow g^{-}(x_{i})$ bounding $i$'s share $y_{i}$ from
below. The function $g^{-}$ is a feasible guarantee if and only if $%
\sum_{1}^{n}g^{-}(x_{i})\leq \mathcal{W}(x)$ for all profiles $x$. If the
output of $\mathcal{W}$ is a cost, an \textit{upper guarantee} $%
x_{i}\rightarrow g^{+}(x_{i})$ bounds $y_{i}$ from above and is feasible iff 
$\mathcal{W}(x)\leq \sum_{1}^{n}g^{-}(x_{i})$ for all $x$.

We take the fair share concept one step further. Given the technology $%
\mathcal{W}$ we look for a \textit{pair} of $(g^{-},g^{+})$ of mutually
compatible guarantees, one lower and one upper. The interval $%
[g^{-}(x_{i}),g^{+}(x_{i})]$ captures all that I stand to gain or lose at
the interim (aka ex ante) stage where I know my type but am clueless about
other agents' types.The smaller the stakes, the less risky to participate,
the less other agents can influence my own welfare. Taking the reduction of
the externalities between types as our normative goal, we are looking for
the smallest (inclusion-wise) correspondences $x_{i}\rightarrow \lbrack
g^{-}(x_{i}),g^{+}(x_{i})]$ meeting the inequalities%
\begin{equation}
\sum_{1}^{n}g^{-}(x_{i})\leq \mathcal{W}(x)\leq \sum_{1}^{n}g^{+}(x_{i})%
\text{ for all profiles }x  \label{32}
\end{equation}

Upper bounding the agents'welfare, although less common than lower bounding
it, appears regularly in the form of \textquotedblleft no
subsidy\textquotedblright\ constraints in the natural monopoly, cost sharing
and cooperative game literatures (\textbf{Fa,} \textbf{Shar, BPW, She }). If
a lower bound on my welfare protects \textit{me} against other agents, an
upper bound protects \textit{them} as a group by limiting the amount of
surplus I can grab. Our examples, starting with the iconic Example 2.1 in
the next section, vindicate the relevance of this viewpoint.

Because $\mathcal{W}$ treats agents symmetrically, our search for the pairs
of tight guarantees $(g^{-},g^{+})$ simplifies to two independent searches
(Lemma 3.5): one for the lower guarantees $x_{i}\rightarrow g^{-}(x_{i})$
that cannot be increased for any type without violating feasibility; and one
for the upper guarantees $x_{i}\rightarrow g^{+}(x_{i})$ that,
symmetrically, cannot be decreased for any type. We call them \textit{tight}
guarantees: they are the closest separably additive symmetric approximations
of $\mathcal{W}$ from above and below.

By construction, given a pair $(g^{-},g^{+})$ of tight guarantees we can
easily define many deterministic sharing rules meeting the inequalities (\ref%
{32}): for instance a simple interpolation will do (Lemma 3.6). Conversely
any deterministic rule dividing the output at all profiles of types and
treating agents symmetrically generates its own pair of guarantees $%
(g^{-},g^{+})$ that may or may not be tight. The extra step to choose a
deterministic sharing rule implementing a given guarantee typically requires
additional context-dependent axiomatic properties. Two notable exceptions
are described three paragraphs below.

\paragraph{The punchlines}

Looking for answers to this elegant mathematical question , we discover the
critical role of the functions $\mathcal{W}$ we call \textit{semi modular},
i.e., supermodular or submodular.\footnote{%
Loosely, for $i\neq j$ the derivative $\partial _{i}\mathcal{W}(x)$
increases weakly in type $x_{j}$ everywhere (or decreases weakly everywhere).%
}

Without any further assumption on $\mathcal{W}$ we can describe
\textquotedblleft more than half\textquotedblright\ of the tight solutions
of system (\ref{32}): on one side there is a simple, unique tight solution
we call the \textit{Unanimity} one (Proposition 4.1); on the other side the
infinite set (typically of infinite dimension) of tight solutions is
bookended by two \textit{Stand Alone} guarantees (Theorem 4.1). We explain
these terms shortly in Section 1.1.

For all two person (strictly) semi-modular functions $\mathcal{W}$ we
characterise the entire set of tight guarantees (Theorem 7.1). We offer no
such result for three or more persons.

To implement the pair with the Unanimity guarantee on one side one of the
Stand Aloneguarantees on the other, we can use the version of serial cost
sharing rule (\textbf{\cite{She}, \cite{MoSh1} \cite{deF}, \cite{AZ}})
suited for this model: Proposition 4.2.

In the classic commons where types are substitute, $\mathcal{W}(x)$ becomes $%
F(\sum_{1}^{n}x_{i})$ we uncover two sequences (one discrete, one
continuous) of tight guarantees connecting the two canonical Stand Alone
ones (Propositions 5.1 and 5.2)\textit{. }We also get a clear view of the
normative consequences of choosing a particular pair of tight guarantees.
For instance if the output is a desirable commodity and $F$ is increasing,
most of these pairs penalise small inputs (that do not pull their weight, so
to speak) and reward large ones much more sharply than the familiar Average
Returns or Shapley Shubik rules.

The proof of our methodology is in the pudding. We apply it to a versatile
set of model in Sections 2, 5 and 6, some well known and some not. The
seemingly neutral search for tight guarantees ends up offering a
parametrised set of interpretations of fairness, often with
\textquotedblleft responsibility flavour\textquotedblright\ :a type is
punished or rewarded for deviating from an exogenous benchmark type,
representing the expected level of effort-input, choice of a location, etc..

\subsection{Overview of the results and examples}

The per capita value of $\mathcal{W}$ at a diagonal profile $(x_{i},\cdots
,x_{i})$ is type $x_{i}$'s \textit{Unanimity} share $una(x_{i})=\frac{1}{n}%
\mathcal{W}(x_{i},\cdots ,x_{i})$. This concept is central to the fair
division literature, as explained in Section 1.2 and confirmed in all our
examples. Applying the inequalities (\ref{32}) to the profile $(x_{i},\cdots
,x_{i})$ we see that \textit{any} lower and upper guarantees satisfy $%
g^{-}(x_{i})\leq una(x_{i})\leq g^{+}(x_{i})$ for all types. Therefore, if $%
una$ happens to be a (lower or upper) guarantee itself, it is the only tight
one (Lemma 3.5).

Prior to the systematic analysis, we give a particularly simple instance of
our approach (Section 2). The famous (submodular) commons $\mathcal{W}%
(x)=\max_{1\leq i\leq n}\{x_{i}\}$ is interpreted by Littlechild and Owen (%
\cite{LO}) as the cost of a capacity shared by agents $i$ with different
needs $x_{i}$. Alternatively a single indivisible good is the common
property of the agents and $x_{i}$ is $i$'s willingness to pay for the good;
the good goes to an efficient agent who must compensate the other owners.
The Unanimity share $\frac{1}{n}x_{i}$ is a guaranteed benefit in the latter
story but a minimal cost share in the former. The tight upper guarantees
form a one dimensional interval parametrised by a type $p$\ interpreted as a
benchmark cost share or price: Lemma 2.1and its comments.

Section 3 introduces tight guarantees \ for a fully general function $%
\mathcal{W}$ and the important notion of contact profile (Lemma 3.2), then
reviews their regularity (monotonicity, continuity) and invariance
properties: Lemmas 3.3 to 3.5. We also define the important class of \textit{%
simple} guarantees parametrised by up to $n-1$ fixed types, and its subset
of \textit{Stand Alone} guarantees captured by a single parameter.

Our first main results are in Section 4, and rest on the super or sub
modularity of $\mathcal{W}$. Because $\mathcal{W}$ is symmetric, it is super
(resp sub) modular if $\mathcal{W}(x_{1}^{\ast };x_{-1})-\mathcal{W}%
(x_{1};x_{-1})$ increases (resp decreases) weakly in the variables $x_{-1}$.
The Unanimity share is the unique tight upper guarantee of $\mathcal{W}$ if
it is supermodular, its unique lower guarantee if it is submodular:
Proposition 4.1: so half of our search is over already. On the other side of
the unanimity the two canonical \textit{Stand Alone} guarantees mentioned
above are $g_{L}$ and $g_{H}$ where $L$ and $H$ are the lower and upper
bounds of the interval of types. For each type $x_{i}$%
\begin{equation*}
g_{L}(x_{i})=\mathcal{W}(x_{i},\overset{n-1}{\overbrace{L,..,L}})-\frac{n-1}{%
n}\mathcal{W}(\overset{n}{\overbrace{L,\cdots ,L}})
\end{equation*}%
\begin{equation*}
g_{H}(x_{i})=\mathcal{W}(x_{i},\overset{n-1}{\overbrace{H,..,H}})-\frac{n-1}{%
n}\mathcal{W}(\overset{n}{\overbrace{H,\cdots ,H}})
\end{equation*}

Note that $g_{L}(L)=una(L)$ and $g_{L}(x_{i})$ adds to $una(L)$ the shift in
output $\mathcal{W}(x_{i},\overbrace{L,..,L})-\mathcal{W}(\overbrace{L,..,L}%
) $ when agent $i$ deviates from the benchmark $L$. Like the Unanimity
share, the guarantees $g_{L}$ and $g_{H}$ appear frequently in the
literature, in particular for the classic commons of Section 5. (see Section
1.2). The same applies to $g_{H}$ at the benchmark $H$.

Theorem 4.1 identifies the bookends\textit{\ of} $\mathcal{G}$ as $g_{L}$
and $g_{H}$. To fix ideas say $\mathcal{W}$ is supermodular. Then every
tight lower guarantee $g^{-}$starts at $L$ above $g_{H}(L)$ and below $%
g_{L}(L)=L$, then ends at $H$ above $g_{L}(H)$ and below $g_{H}(H)=H$. Also, 
$g$ grows slower than $g_{H}$\ and faster than $g_{L}$ (a weak
single-crossing property).

Finally we adapt to our model the two serial cost sharing rules, one
starting from $L$ the other from $H$ (\cite{MoSh1}, \cite{deF}). They
implement respectively the tight pair $(una,g_{L})$ and $(una,g_{H})$:
Proposition 4.2.\smallskip

We turn in Section 5 to the classic commons where the inputs are perfect
substitutes: $\mathcal{W}(x)=F(x_{N})$ (where $x_{N}$ stands for $%
\sum_{1}^{n}x_{i}$). It is super (sub) modular if and only if $F$ is convex
(concave).

Of the huge family of tight guarantees on the other side of the Unanimity
share, \ that we do not describe, we identify two subsets connecting $g_{L}$
and $g_{H}$. First the sequence of simple guarantees $g_{\ell
,h}(x)=F(x_{i}+\ell L+hH)-C$ where $C$ is a constant and $\ell ,h$ are two
integers summing up to $n-1$ (so $g_{L}$ obtains for $\ell =n-1$ and $g_{H}$
for $h=n-1$): Proposition 5.1. Next a continuous one dimensional family made
mostly of tangents to the graph of the function $x_{i}\rightarrow una(x_{i})$%
: Proposition5.2. Three examples illustrate their compromising role between $%
g_{L}$ and $g_{H}$: sharing the cost of a public bad (Example 5.1); a
commons with complementary inputs (Example 5.2); and sharing the variance of
types on a line (Example 5.3).

Section 6 introduces the finite dimensional class of \textit{rank-separable}
functions: they are separably additive with respect to the order statistics
of types. The simplest example is $\mathcal{W}(x)=\max_{i}\{x_{i}\}$
(Section 2). If a rank-separable function is semi-modular as well (like $%
\max_{i}\{x_{i}\}-\min_{i}\{x_{i}\}$, but not $\max_{i}\{x_{i}\}+\min_{i}%
\{x_{i}\}$), its tight guarantees opposite to the Unanimity are exactly all
the simple guarantees in Definition 3.3. Conversely, among semi-modular
functions\textbf{\ }the rank-separable ones are characterised by this
property: Theorem 6.1. Three examples include a familiar queuing model:
Example 6.1; sharing the cost of a spread: Example 6.2; and our single
example of a \textit{non }semi-modular function: Example 6.3.

Theorem 7.1 in Section 7 is a full characterisation of \textit{all} tight
lower guarantees for two person problems when $\mathcal{W}(x_{1},x_{2})$ is 
\textit{strictly} supermodular and, symetrically, if $\mathcal{W}$ is
strictly submodular. They are parametrised by the choice of a decreasing,
continuous and symmetric function from the set of types into itself. So the
full set of tight guarantees is much larger, in fact of infinite dimension,
when the semi-modularity property is strict. This contrast with the (not
strictly semi-modular) rank separable functions above, of which the tight
guarantees are parametrised by $[L,H]^{n-1}$.

After listing some open questions in Section 8, the Appendix (Section 9)
contains many long or minor proofs.

\subsection{Related literature}

\paragraph{Sharing private commodities}

In the original cake cutting model individual utilities are additive and non
atomic over the cake. At a unanimous profile of utilities every agent $i$
can receive a share worth $\frac{1}{n}$ of the whole cake; the corresponding
"Proportionalty" property is precisely the Unanimity tight lower guarantee (%
\cite{St}, \cite{Ku} \cite{DS}). For the allocation of a divisible private \
goods \`{a} la Arrow Debreu when individual preferences are convex, agent $i$%
's Unanimity utility is $u_{i}(\frac{1}{n}\omega )$, again the compelling
endogenous fair share of that model (\textbf{\cite{Va}}). If we allocate
indivisible goods and use cash transfers to smoothen individual utilities,
the Unanimity lower guarantee is the only fair share that fits non
interpersonally comparable ordinal preferences (\cite{TT}, \cite{ADG}).

However, if utilities over the private cake are not additive, or if
preferences for Arrow Debreu commodities are not necessarily convex, the
Unanimity utility can be defined but is no longer a guarantee from below or
from above. In such cases almost nothing is known about tight guarantees (%
\cite{BM}).

In the 21st century a considerable literature is still looking for a
compelling definition of the fair share when we allocate indivisible items
(good or bad),utilities are additive, but no monetary compensations or
randomisation is feasible. The plausible \textit{MaxMinShare} utility (\cite%
{Bu}) is the correct interpretation of the Unanimity utility. But only a
little more than $\frac{3}{4}$ of this share can be guarantee (\cite{PW}, 
\cite{AG}). The search for a convincing concept of guaranteed utility is
still open. Various proposals include \cite{DH}, \cite{MaPs}, \cite{LMSZ}
and \cite{BaFe}.

\paragraph{Cooperative production}

To our bare bones model of the commons we can add individual preferences
over the bundles $(x_{i},y_{i})$ (a share of input and one of output). Then
the Stand Alone welfare comes from using a private copy of the production
function, and the Unanimity welfare from sharing it with $n-1$ equal agents
who share my preferences. When marginal returns are either increasing or
decreasing, the SA welfare stands on opposite side of the Pareto frontier
from the Unanimity one (\cite{Sha}, \cite{FrMo}, \cite{Mo4}, \cite{Wa}). The
same is true in the public good provision model irrespective of returns:
Standing Alone means choosing the public good level and paying its full cost
-- worst case --, Unanimity means choosing this level and paying $\frac{1}{n}
$-th of its cost (\cite{Mo4}\textbf{)}.

The \textit{undirected} Unanimity test (\cite{Mo6}, \cite{Th}, \cite{Wa})
uses the profile of Unanimity utilities to either lower or upper bound
individual utilities, whichever is feasible for a particular profile of
types. It is meaningful in the cooperative production context where variable
returns imply that this profile can easily change side.

\paragraph{Mathematical literature}

Although the question of approximating a real valued function $%
f(x_{1},\cdots ,x_{n})$ by separably additive functions from above and below
is mathematically very natural, we could only find one paper (\cite{SW})
addressing it and this, perhaps unsurprisingly, in the economics literature.
Theorem 2 there proves statement $i)$ in our Theorem 4.1 without assuming
that $f$ is symmetric in the $x_{i}$-s, a critical assumption for most of
our other results.

A formal connection\footnote{%
We thank Fedor Sandomirskiy for pointing it out.} of our approximation
question to the celebrated Optimal Transport problem (\cite{Vi}, \cite{Ga})
comes from its dual formulation as the Kantorovitch- Rubinstein Lemma:%
\newline
$\max_{\Pi :\Pi _{i}=\lambda _{i}}\{\int \mathcal{W}(x)d\Pi
(x)\}=\min_{g_{i}:\sum_{i}g_{i}(x_{i})\geq \mathcal{W}(x)}\{\sum_{i}\int
g_{i}(x_{i})d\lambda _{i}\}$\newline
where $\mathcal{W}(x)$ is the abstract transport cost, and $\Pi $ the
transportation protocol with fixed marginals $\lambda _{i}$ over the $n$
coordinates of $x$. Again this literature does not assume that $\mathcal{W}$
is symmetric, and neither does its economic applications to matching (\cite%
{Ga}).

\section{Example 2.1}

The generic type $x_{i}$ varies in the interval $[0,H]$ and $\mathcal{W}%
(x)=\max_{1\leq i\leq n}\{x_{i}\}$.

\textbf{Story 1: }\textit{The agents own an indivisible good and }$x_{i}$%
\textit{\ is }$i$\textit{'s willingness to pay for it. One of the efficient
agents will get it and compensate the others in cash. What compensations are
fair?}

\textbf{Story 2:} \textit{They share the (1 to 1) cost of a public capacity:
the length of a runway (\cite{LO}), the width of a channel, or the computing
power of a machine they share. Type }$x_{i}$\textit{\ is the cost of the
capacity agent }$i$\textit{\ needs. Who should pay how much?}

We use simple division rules to generate some lower and upper guarantees,
automatically feasible but not necessarily tight. In story 1 we could treat
the efficient surplus as a common property and split it equally: $y_{i}=%
\frac{1}{n}\mathcal{W}(x)$. This rule generates the following guarantees%
\begin{equation*}
g^{-}(x_{i})=\min_{x_{-i}}\frac{1}{n}\max \{x_{i},\max_{j\neq i}x_{j}\}=%
\frac{1}{n}x_{i}=una(x_{i})
\end{equation*}%
\begin{equation*}
g_{H}^{+}(x_{i})=\max_{x_{-i}}\{\frac{1}{n}\max \{x_{i},\max_{j\neq
i}x_{j}\}\}=\frac{1}{n}H
\end{equation*}

The Unanimity guarantee is the uncontroversial unique tight lower guarantee:
I am entitled to a fair share of the surplus I generate by consuming the
good. Check now that\ the upper bound $g_{H}^{+}$ is tight. Fix $x_{i}$ and
use the notation $(x_{i},\overset{n-1}{H})$ for a profile where $n-1$
coordinates are $H$. At such profile all agents including $i$ cannot get
more than $\frac{1}{n}H$ but we distribute $H$ so $y_{i}=\frac{1}{n}H$: so
the upper bound $g_{H}^{+}(x_{i})$ cannot be lowered at $x_{i}$.

In story 2 the share $\frac{1}{n}x_{i}$ is compelling as a minimal payment
by $i$ ($i$'s best case), but the upper guarantee $\frac{1}{n}H$ ($i$'s
worst case) less so: why would an agent with small needs pay as much as all
others who ask for the largest capacity $H$? It seems more natural to divide
total cost in proportion to individual needs. That rule implements the
guarantees%
\begin{equation*}
g^{-}(x_{i})=\min_{x_{-i}}\{\frac{x_{i}}{x_{i}+x_{N\diagdown i}}\max_{1\leq
j\leq n}\{x_{j}\}\}=\frac{1}{n}x_{i}=una(x_{i})
\end{equation*}%
\begin{equation*}
g_{0}^{+}(x_{i})=\max_{x_{-i}}\{\frac{x_{i}}{x_{i}+x_{N\diagdown i}}%
\max_{1\leq j\leq n}\{x_{j}\}\}=x_{i}
\end{equation*}

The upper guarantee $g_{0}^{+}$ is the familiar Stand Alone cost (worst
case). It is tight: at the profile $(x_{1},\overset{n-1}{0})$ no agent other
than $1$ pays anything (because $una(0)=g_{0}^{+}(0)=0$), therefore $i$
covers the full cost.

In story 1 the upper guarantee $g_{0}^{+}(x_{i})=x_{i}$ (best case) makes
sense if the good is worthless outside the $n$ agents (e. g., a heirloom
with only sentimental value): I am not entitled to more surplus than what I
generate by eating the good for free.

Clearly all strict convex combinations of $g_{0}^{+}$ and $g_{H}^{+}$ are
upper guarantee of $\mathcal{W}$. Our first result implies that none of them
is tight.\smallskip

\textbf{Lemma 2.1:}\textit{\ }We use the notation $z_{+}=\max \{z,0\}$.%
\newline
\textit{The tight \textit{upper guarantee}s of} $\mathcal{W}(x)=\max_{1\leq
i\leq n}\{x_{i}\}$\textit{\ are parametrised by a type }$p\in \lbrack 0,H]\ $%
\textit{as} $g_{p}^{+}(x_{i})=\frac{1}{n}p+(x_{i}-p)_{+}$ for $x_{i}\in
\lbrack 0,H]$.\textit{\textit{\smallskip }}

So $g_{0}^{+}$ and $g_{H}^{+}$ are the two end-points of the interval of
tight upper guarantees, a special case of Proposition 4.2.

The Lemma follows from the (much) more general Theorem 6.1. The following
simple proof gives some intuition for our techniques.

\textbf{Proof}: Checking the inequality $\sum_{1}^{n}g_{p}^{+}(x_{i})\geq
\max_{1\leq i\leq n}\{x_{i}\}$ is routine. Next we pick an arbitrary tight
upper guarantee $g^{+}$ of $\mathcal{W}$ and set $p=ng^{+}(0)$. At the
unanimous profile $\overset{n}{(0})$ the inequality (\ref{32}) implies $%
p\geq 0$. Tightness implies that $g^{+}$ increases weakly (this is easy to
check or see Lemma 3.1), so $g^{+}(x_{i})\geq \frac{1}{n}p$ for all $x_{i}$.
The constant function $\frac{1}{n}H$ is an upper guarantee therefore if $p>H$
the guarantee $g^{+}$ is not tight. So $p\in \lbrack 0,H]$. Applying again (%
\ref{32}) to $(x_{i},\overset{n-1}{0})$ gives $g^{+}(x_{i})\geq x_{i}-\frac{%
n-1}{n}p$. Combining this with $g^{+}(x_{i})\geq \frac{1}{n}p$ we get $%
g^{+}\geq g_{p}^{+}$. Because $g^{+}$ is tight and $g_{p}^{+}$ is an upper
guarantee this must be an equality and $g_{p}^{+}$ is tight too. $%
\blacksquare \smallskip $

Choosing the tight pair $(una,g_{p}^{+})$ implies that the benchmark type $p$
gets the share $\frac{1}{n}p$ \textit{irrespective of other agents' types}.
Story 1: $p$ could be an estimate of the market value of the good. If the
good is worth less than $p$ to $i$, she will receive \textit{at most} a fair
share of $p$; if $i$ values it above $p$ she \textit{could} receive, in
addition to $\frac{1}{n}p$ the full surplus $x_{i}-p$. This will happen for
sure if $x_{i}$ is the only type above $p$: then $i$ gets the good and pays $%
\frac{1}{n}p$ to everyone else.

Story 2: $p$ could be the normal (e. g., status quo) capacity; only agents
with needs larger than $p$ can be charged more than $\frac{1}{n}p$, and the
surcharge can reach the full incremental cost $x_{i}-p$. Overall an agent
with small needs prefer a low benchmark capacity $p$ and one with large
needs a high $p$. And vice versa in story 1 where the agents with a large
type like the benchmark price $p$ to be low.

\textbf{Figure 1 }\textit{shows, in a three agent example, the tight
guarantees }$una$\textit{, }$g_{0}^{+}$\textit{, }$g_{H}^{+}$\textit{\ and }$%
g_{p}^{+}$\textit{.}

\paragraph{sharing a chore and cash compensation}

Consider now the function $\mathcal{W}^{\bigstar }(x)=\min_{i\in N}\{x_{i}\}$
describing the assignment of an indivisible chore (a bad). For efficiency an
agent $i^{\ast }$ with minimal disutility $x_{i^{\ast }}$ does the chore and
is compensated (with cash) by the other agents. The Unanimity share is still 
$\frac{1}{n}x_{i}$ but now the unique tight upper guarantee: I don't agree
to work more than \textquotedblleft my fair share\textquotedblright\ of the
chore.

The function $\mathcal{W}^{\bigstar }$ obtains from $\mathcal{W}$ by the
(bijective) change of variable $x_{i}=-z_{i}$ so we can apply statement $ii)$
in Lemma 3.4 and deduce the tight \textit{lower} guarantees $g^{\bigstar -}$
of $\mathcal{W}^{\bigstar }$ from the tight upper ones of $\mathcal{W}$. \
They are also parametrised by a benchmark disutility $p\in \lbrack 0,H]$ as $%
g_{p}^{\bigstar -}(x_{i})=\frac{1}{n}p+(x_{i}-p)_{-}$ with the notation $%
(z)_{-}=\min \{z,0\}$.

So an agent who dislike the chore more than $p$ must pay at least $\frac{1}{n%
}p$. And one with type $x_{i}$ below $\frac{n-1}{n}p$ \textit{may} end up
with a net profit (paid more than his disutility) for doing the chore. At
the lower guarantee $g_{H}^{\bigstar -}(x_{i})=x_{i}-\frac{n-1}{n}H$ the
efficient agent may extract $\frac{1}{n}H$ from every other agent if they
dislike the chore at the level $H$.

\section{Guarantees: definition and general properties}

\textbf{Definition 3.0} \textit{A commons problem is a triple} $(N,\mathcal{X%
},\mathcal{W})$\textit{\ where }$N$\textit{\ is the set of }$n$\textit{\
agents}, $\mathcal{X}=[L,H]\subset 
\mathbb{R}
$ \textit{is the interval of types, and the real valued function} $\mathcal{W%
}:$ $\mathcal{X}^{N}\ni x\rightarrow \mathcal{W}(x)$ \textit{is symmetric in
the }$n$\textit{\ variables }$x_{i}$\textit{\ and continuous on} $\mathcal{X}%
^{N}$.

\textit{At the profile} $x=(x_{i})_{i\in N}\in \mathcal{X}^{N}$ \textit{we
must divide the benefit or cost }$\mathcal{W}(x)$.\smallskip

The notation $x_{i}$\ can be the type of a specific agent identified by the
context, or simply a generic single type.\smallskip

\textbf{Definition 3.1} \textit{The real valued function }$g^{-}$ (\textit{%
resp }$g^{+}$) \textit{on }$\mathcal{X}$\textit{\ is a lower (resp upper)
guarantee of} $\mathcal{W}$ \textit{if it satisfies} \textit{the following
Left (resp Right) Hand inequality:} 
\begin{equation}
\sum_{i\in N}g^{-}(x_{i})\leq \mathcal{W}(x)\leq \sum_{i\in N}g^{+}(x_{i})%
\text{ for all }x\in \mathcal{X}^{N}  \label{3}
\end{equation}

\textit{We write} $\boldsymbol{G}^{-},\boldsymbol{G}^{+}$\textit{\ for the
(clearly non empty) sets of such guarantees.\smallskip }

\textbf{Definition 3.2 }\textit{Given two lower guarantees }$%
g_{1}^{-},g_{2}^{-}\in G^{-}$\textit{\ we say that }$g_{1}^{-}$\textit{\
dominates }$g_{2}^{-}$\textit{\ if }$g_{1}^{-}(x_{i})\geq g_{2}^{-}(x_{i})$%
\textit{\ for }$x_{i}\in \mathcal{X}$ \textit{and }$g_{1}^{-}\neq g_{2}^{-}$%
\textit{. The guarantee }$g^{-}\in G^{-}$\textit{\ is tight if it is not
dominated in }$G^{-}$; \textit{equivalently increasing }$g^{-}$\textit{\ at
a single} $x_{1}\in \mathcal{X}$ \textit{creates a violation of the LH
inequality in (\ref{3}) for some} $x_{-1}\in \mathcal{X}^{[n-1]}$.

\textit{The isomorphic statement for upper guarantees in }$G^{+}$\textit{\
flips the domination inequality around; for tightness it replaces increasing
by decreasing and LH by RH.}

\textit{We write} $\mathcal{G}^{-}$ \textit{and} $\mathcal{G}^{+}$ \textit{%
for the subsets of tight guarantees in} $\boldsymbol{G}^{-}$ \textit{and} $%
\boldsymbol{G}^{+}$.\smallskip

\textbf{Lemma 3.1 }$i)$ \textit{For }$\varepsilon =+,-$ a\textit{\ guarantee 
}$g\in \boldsymbol{G}^{\varepsilon }$ \textit{is either tight or dominated
by a tight one. So }$\mathcal{G}^{\varepsilon }$ \textit{is not empty}.

\noindent $ii)$ \textit{A tight guarantee is weakly monotonic in }$x_{i}$ 
\textit{if }$\mathcal{W}$ \textit{is in }$x$\textit{.}

\noindent $iii)$ \textit{A tight guarantee is continuous in }$x_{i}$ \textit{%
if }$\mathcal{W}$\textit{\ is in }$x$\textit{.\smallskip }

Statement $i)$ is a simple application of Zorn's Lemma. For the monotonicity
statement in $ii)$ fix $g\in \mathcal{G}^{-}$. If $x_{i}>x_{i}^{\ast }$ and $%
g(x_{i})<g(x_{i}^{\ast })$ define $\widetilde{g}(x_{i})=g(x_{i}^{\ast })$
and $\widetilde{g}=g$ otherwise, then check that $\widetilde{g}$ is still in 
$\boldsymbol{G}^{-}$. This contradicts that $g$ is tight.\ The longer proof
of continuity is in Section 9.1.\smallskip

\textbf{Lemma 3.2 }\textit{A guarantee }$g$\textit{\ in }$\boldsymbol{G}%
^{\varepsilon }$\textit{\ is tight if and only if for all }$x_{i}\in 
\mathcal{X}$ \textit{there exists} $x_{-i}\in \mathcal{X}^{[n-1]}$ \textit{%
such that }$g(x_{i})+\sum_{j\neq i}g(x_{j})=\mathcal{W}(x_{i},x_{-i})$. 
\textit{Then we call }$(x_{i},x_{-i})$\textit{\ a contact profile of }$g$ 
\textit{at }$x_{i}$. \textit{The set of such profiles\ is the contact set }$%
\mathcal{C}(g)$\textit{\ of }$g$.

Statement "if" is clear. The proof of "only if" is in Section 9.1\textbf{%
.\smallskip }

\textbf{Lemma 3.3 }\textit{Fix }$g\in \mathcal{G}^{+}$. \textit{For any }$%
x_{i},x_{i}^{\ast }$\textit{\ and contact profile }$x=(x_{i},x_{-i})$ 
\textit{of }$g$\textit{\ at }$x_{i}$ \textit{we have }$g(x_{i}^{\ast
})-g(x_{i})\geq \mathcal{W}(x_{i}^{\ast },x_{-i})-\mathcal{W}(x_{i},x_{-i})$ 
\textit{and the opposite inequality if }$g\in \mathcal{G}^{-}$.\smallskip

\textbf{Proof }In the inequality $g(x_{i}^{\ast })+\sum_{j\neq
i}g(x_{j})\geq \mathcal{W}(x_{i}^{\ast },x_{-i})$ we replace each term $%
g(x_{j})$ by $\mathcal{W}(x)-g(x_{i})-\sum_{k\neq i,j}g(x_{k})$ and get $%
(n-1)(\mathcal{W}(x)-g(x_{i}))-(n-2)\sum_{j\neq i}g(x_{j})\geq \mathcal{W}%
(x_{i}^{\ast },x_{-i})-g(x_{i}^{\ast })$. Equivalently $\mathcal{W}%
(x)-g(x_{i})+(n-2)(\mathcal{W}(x)-\sum_{N}g(x_{j}))\geq \mathcal{W}%
(x_{i}^{\ast },x_{-i})-g(x_{i}^{\ast })$. The term in parenthesis is zero by
our choice of $x_{-i}$ so we are done. $\blacksquare \smallskip $

\textbf{Corollary} \textit{Fix }$K$\textit{\ positive such that the function}
$\mathcal{W}$ \textit{is }$K$\textit{-Lipschitz in }$x_{i}$\textit{,
uniformly in} $x_{-i}\in \mathcal{X}^{[n-1]}$\textit{. Then so is each} 
\textit{tight guarantee} $g\in \mathcal{G}^{\varepsilon }$ for $\varepsilon
=+,-$.\smallskip

We state without the simple proof two useful invariance properties of
guarantees.\smallskip

\textbf{Lemma 3.4 }\textit{Fix}\textbf{\ }$\mathcal{X}=[L,H]$\textit{, }$%
\mathcal{W}$ \textit{and} $\varepsilon =+,-$.

\noindent $i)$\textit{\ If} $\mathcal{W}_{0}$ \textit{is} \textit{additively
separable, }$\mathcal{W}_{0}(x)=\sum_{N}w_{0}(x_{i})$,\textit{\ then }$%
\mathcal{G}^{\varepsilon }(\mathcal{W}+\mathcal{W}_{0})=\mathcal{G}%
^{\varepsilon }(\mathcal{W)}+\{w_{0}\}$

\noindent $ii)$\textit{\ Fix }$\theta $\textit{\ a bicontinuous increasing
bijection }$x_{i}=\theta (z_{i})$\textit{\ from }$\mathcal{Z}=[\theta
^{-1}(L),\theta ^{-1}(H)]$ \textit{into} $\mathcal{X}$, \textit{and change
variables to a new problem }$\mathcal{W}^{\blacklozenge }(z)=\mathcal{W}%
(\theta (z))$ \textit{where }$\theta (z)_{i}=\theta (z_{i})$\textit{\ for
all }$i$. \textit{If }$g\in \mathcal{G}^{\varepsilon }(\mathcal{W)}$\textit{%
\ then }$g\circ \theta \in \mathcal{G}^{\varepsilon }(\mathcal{W}%
^{\blacklozenge }\mathcal{)}$\textit{. If }$\theta \ $\textit{is decreasing,
ceteris paribus, then} $g\circ \theta \in \mathcal{G}^{-\varepsilon }(%
\mathcal{W}^{\blacklozenge }\mathcal{)}$.\textit{\smallskip }

\subsection{Unanimity shares and guarantees}

\textit{Notation}: we write $(\overset{k}{y};\overset{k^{\prime }}{z})$\ for
any $(k+k^{\prime })$-vector where $k$ coordinates are $y$\ and $k^{\prime }$%
\ are $z$. As $\mathcal{W}$\ is symmetric in its variables, the expression $%
\mathcal{W}(\overset{k}{y};\overset{k^{\prime }}{z})$ is well-defined.

The function $\mathcal{W}$ on the diagonal of $\mathcal{X}^{N}$ defines the 
\textit{Unanimity }share $una(x_{i})=\frac{1}{n}\mathcal{W}(\overset{n}{x_{i}%
})$.\smallskip

\textbf{Lemma 3.5 }$i)$ \textit{For any }$x_{i}\in \mathcal{X}$\textit{\ and 
}$(g^{-},g^{+})\in \boldsymbol{G}^{-}\times \boldsymbol{G}^{+}$ \textit{we
have }$g^{-}(x_{i})\leq una(x_{i})\leq g^{+}(x_{i})$.

\noindent $ii)$ \textit{For }$\varepsilon =+,-$ \textit{if }$una\in 
\boldsymbol{G}^{\varepsilon }$\textit{\ it dominates every other guarantee
in }$\boldsymbol{G}^{\varepsilon }$ \textit{and} $\mathcal{G}^{\varepsilon
}=\{una\}$.

\textit{Conversely,} \textit{if }$\mathcal{G}^{\varepsilon }$ \textit{is a
singleton it must be the Unanimity guarantee:} $|\mathcal{G}^{\varepsilon
}|=1\Longrightarrow \mathcal{G}^{\varepsilon }=\{una\}$.

\noindent $iii)$ \textit{For }$\varepsilon =+,-$\textit{\ and any} $x_{i}\in 
\mathcal{X}$ \textit{there is a tight guarantee} $g\in \mathcal{G}%
^{\varepsilon }$ \textit{s.t}. $g(x_{i})=una(x_{i})$.\smallskip

Statement $i)$ and the first part of $ii)$ follow at once by applying
inequalities (\ref{3}) at a unanimous profile. The rest of the proof is in
Section 9.2.\smallskip

Statement $i)$ implies at once the decoupling property announced in Section
1 (two paragraphs before \textit{The punch lines}). Fix a correspondence $%
x_{i}\rightarrow J(x_{i})$ from types to closed real intervals s. t. at any
profile $x$ each $i$ can get a share in $J(x_{i})$ while dividing $\mathcal{W%
}(x)$. Then $J$ is inclusion minimal with this property if and only if $%
J(x_{i})\equiv \lbrack g^{-}(x_{i}),g^{+}(x_{i})]$ for some pair $%
(g^{-},g^{+})\in $ $\mathcal{G}^{-}\times \mathcal{G}^{+}$.

Finally if both sets $\mathcal{G}^{\varepsilon }$ are singletons statement $%
ii)$ implies that $una$ meets both inequalities in (\ref{3}) therefore $%
\mathcal{W}(x)=\sum_{i=1}^{n}una(x_{i})$ is an additively separable function.

\subsection{A special class of guarantees}

In Example 2.1, for any $x_{i}$ one of the contact profiles of $g_{p}$ at $%
x_{i}$ (Lemma 3.2) is $(x_{i},\overset{n-1}{p})$. The guarantees for which a
fixed $(n-1)$-profile serves as contact profile for all types are easy to
describe and use; they play a key role in Sections 4,5 and 6.\smallskip

\textbf{Definition 3.3 }\textit{An upper or lower guarantee is called simple
and denoted }$g_{c}$ \textit{if }$(x_{i},c)$ \textit{is a contact profile of 
}$g_{c}$\textit{\ for all }$x_{i}\in \mathcal{X}$. That is $%
g_{c}(x_{i})+\sum_{\ell =1}^{n-1}g_{c}(c_{\ell })=\mathcal{W}(x_{i},c)$.

\textit{Equivalently }$g_{c}$\textit{\ is defined as} 
\begin{equation}
g_{c}(x_{i})=\mathcal{W}(x_{i},c)-\frac{1}{n}(\sum_{\ell =1}^{n-1}\mathcal{W}%
(c_{\ell },c))\text{ for all }x_{i}\in \mathcal{X}  \label{22}
\end{equation}

To check this equivalence\textbf{\ }note that the contact property of $g_{c}$
implies $g_{c}(x_{i})=\mathcal{W}(x_{i},c)-C$ for some constant $C$. Replace
accordingly each term $g_{c}(c_{\ell })$ by $\mathcal{W}(c_{\ell },c)-C$ and
rearrange to find equation (\ref{22}). The converse move from (\ref{22}) to
the contact property is just as easy.\smallskip

\textbf{Remark 3.1} The following consequence of Lemma 3.2 is useful in the
proofs of Sections 9.4, 9.7 and 9.10.1. For any $c\in \mathcal{X}^{n-1}$ if $%
g_{c}$\ given by (\ref{22}) is a guarantee, it is tight.\textit{\smallskip }

\textbf{Definition 3.4 }\textit{A Stand Alone (SA) guarantee }$g_{c_{0}}$ 
\textit{of} $\mathcal{W}$ i\textit{s a simple guarantee such that }$c=(%
\overset{n-1}{c_{0}})$. It takes the form $g_{c_{0}}(x_{i})=\mathcal{W}%
(x_{i},\overset{n-1}{c_{0}})-\frac{n-1}{n}\mathcal{W}(\overset{n}{c_{0}})$%
.\smallskip

All tight upper guarantees $g_{p}$ in Example 2.1 are Stand Alone guarantees.

The SA guarantee $g_{c_{0}}$ touches\ the Unanimity guarantee at $c_{0}$: $%
g_{c_{0}}(c_{0})=una(c_{0})$. Typically this is not the case for a simple
but not Stand Alone guarantee. See Proposition 5.1.

\subsection{Implementing a guarantee by a sharing rule}

A sharing rule $\varphi $ for $\mathcal{W}$ maps each $x\in $ $\mathcal{X}%
^{N}$ to a division $y=\varphi (x)$ of $\mathcal{W}(x)$: $\sum_{N}y_{i}=%
\mathcal{W}(x)$. Given a sharing rule, it may not be easy to decide whether
the guarantees it implements are tight. Conversely, given a pair $%
(g^{-},g^{+})$ of tight guarantees it is easy to define a rule $\varphi $ of
which the guarantees are precisely this pair.\smallskip

\textbf{Lemma 3.6. }\textit{Fix the function }$\mathcal{W}$\textit{\ and a
pair of tight guarantees }$(g^{-},g^{+})\in \mathcal{G}^{-}\times \mathcal{G}%
^{+}$\textit{. If the sharing rule }$\varphi $\textit{\ satisfies }$%
g^{-}(x_{i})\leq \varphi _{i}(x)\leq g^{+}(x_{i})$\textit{\ for all }$i$%
\textit{\ and }$x$\textit{\ then it implements }$(g^{-},g^{+})$\textit{: for
all }$i$\textit{\ and }$x$\textit{\ we have }$\min_{x_{-i}\in \mathcal{X}%
^{n-1}}\{\varphi _{i}(x_{i},x_{-i})\}=g^{-}(x_{i})$\textit{\ and }$%
\max_{x_{-i}\in \mathcal{X}^{n-1}}\{\varphi
_{i}(x_{i},x_{-i})\}=g^{+}(x_{i}) $.\smallskip

To check the left equality note that $h(x_{i})=\min_{x_{-i}}\{\varphi
_{i}(x_{i},x_{-i})\}$ is a lower guarantee, and is bounded below by $g^{-%
\text{. }}$then invoke the tightness of $g^{-}$.

The moving average of $g^{-}$ and $g^{+}$ is the simplest sharing rule
implementing this pair in $\mathcal{G}^{-}\times \mathcal{G}^{+}$: $\varphi
_{i}(x)=\lambda g^{-}(x_{i})+(1-\lambda )g^{+}(x_{i})$ where for all $x\in 
\mathcal{X}^{N}$ we choose $\lambda $ s. t. $\sum_{N}\varphi _{i}(x)=%
\mathcal{W}(x)$.

\section{Semi-modular functions $\mathcal{W}$}

All our results happen in this class of benefit and cost functions with the
single exception of Example 6.3. The following definitions take into account
that $\mathcal{W}$ is symmetric in its variables.\smallskip

\textbf{Definition 4.1 }\textit{The function }$\mathcal{W}$ is \textit{%
supermodular if for all }$x_{1},x_{1}^{\ast },x_{2},x_{2}^{\ast }$\textit{\
in }$\mathcal{X}$\textit{\ and }$x_{-12}$\textit{\ in} $\mathcal{X}^{n-2}$ 
\textit{such that }$x_{i}\leq x_{i}^{\ast }$\textit{\ for }$i=1,2$\textit{\
we have}%
\begin{equation}
\mathcal{W}(x_{1},x_{2}^{\ast },x_{-12})+\mathcal{W}(x_{1}^{\ast
},x_{2},x_{-12})\leq \mathcal{W}(x_{1},x_{2},x_{-12})+\mathcal{W}%
(x_{1}^{\ast },x_{2}^{\ast },x_{-12})  \label{38}
\end{equation}

\textit{Equivalently for all }$x_{1},x_{1}^{\ast },$\textit{in }$\mathcal{X}$
\textit{such that }$x_{1}<x_{1}^{\ast }$ \textit{the function }$\mathcal{W}%
(x_{1}^{\ast },x_{-1})-\mathcal{W}(x_{1},x_{-1})$ \textit{is weakly
increasing in} $x_{-1}\in \mathcal{X}^{n-1}$. \textit{And} $\mathcal{W}$ 
\textit{is strictly supermodular if this monotonicity is strict.}

\textit{We call }$\mathcal{W}$ \textit{sub-modular (resp. strictly so) if
the opposite inequality of (\ref{38}) holds under the same premises; or if }$%
\mathcal{W}(x_{1}^{\ast },x_{-1})-\mathcal{W}(x_{1},x_{-1})$ \textit{has the
opposite monotonicity properties.}

\textit{We say that }$\mathcal{W}$ is \textit{semi-modular if it is
supermodular or submodular.\smallskip }

Semi-modularity does not require $\mathcal{W}$ to be monotonic in $x$, or
convex or concave even in a single variable. If $\mathcal{W}$ is twice
differentiable semi-modularity means that the sign of $\partial _{ij}%
\mathcal{W}(x)$ is constant sign in $\mathcal{X}^{N}$: $\partial _{ij}%
\mathcal{W}(x)$ is non negative resp non positive for super or submodularity.

The function $\max_{i}\{x_{i}\}$ (Example 2.1) is submodular and $%
\min_{i}\{x_{i}\}$ is supermodular. But for the median coordinate of $%
\{x_{i}\}_{i\in N}$ is not semi-modular: this is easy to check directly, or
see Lemma 6.1.

\subsection{The\ Unanimity guarantee}

\textbf{Proposition 4.1 }\textit{If} $\mathcal{W}$ \textit{is} \textit{%
super, resp sub, modular the Unanimity function }$una(x_{i})=\frac{1}{n}%
\mathcal{W}(\overset{n}{x_{i}})$\textit{\ is its unique tight upper, resp
lower guarantee}.\smallskip

\noindent \textbf{Proof }We assume without loss that $\mathcal{W}$ is
supermodular. Suppose $n=2$. By Lemma 3.5 we only need to check that $una$
is an upper guarantee: $\mathcal{W}(x_{1},x_{2})\leq \frac{1}{2}(\mathcal{W}%
(x_{1},x_{1})+\mathcal{W}(x_{2},x_{2}))$ for any $x_{1},x_{2}$. This follows
from (\ref{38}) and the symmetry of $\mathcal{W}$.

We assume now that the statement is true up to $(n-1)$ agents. We fix a $n$%
-person supermodular function $\mathcal{W}$ and a profile $x\in \mathcal{X}%
^{N}$. By the inductive assumption for all $i$ the Unanimity function of the
supermodular $(n-1)$-function $\mathcal{W(\cdot };x_{i})$ is an upper
guarantee, therefore for all $i$ we have $\mathcal{W(}x)\leq \frac{1}{n-1}%
\sum_{j\in N\diagdown \{i\}}\mathcal{W}(x_{i};\overset{n-1}{x_{j}})$.
Summing up over $i$ gives $n\mathcal{W(}x)\leq \frac{1}{n-1}\sum_{(i,j)\in P}%
\mathcal{W}(x_{i};\overset{n-1}{x_{j}})$, where $P$ is the set of ordered
pairs $(i,j)$ in $N$ (pairs with two distinct components). We set $%
S=\sum_{(i,j)\in P}\mathcal{W}(x_{i};\overset{n-1}{x_{j}})$ and record the
latter inequality as $n\mathcal{W(}x)\leq \frac{1}{n-1}S$.

Fix now a pair $i,j$ and apply the inductive assumption to $\mathcal{W(\cdot 
};x_{j})$ at the $(n-1)$-profile $(x_{i},\overset{n-2}{x_{j}})$: $\mathcal{W(%
}x_{i};\overset{n-1}{x_{j}})\leq \frac{1}{n-1}((n-2)\mathcal{W}(\overset{n}{%
x_{j}})+\mathcal{W(}x_{j};\overset{n-1}{x_{i}}))$. Summing up over all $%
(i,j)\in P$ gives $S\leq (n-2)\sum_{j=1}^{n}\mathcal{W}(\overset{n}{x_{j}})+%
\frac{1}{n-1}S$, simplifying to $S\leq (n-1)\sum_{j=1}^{n}\mathcal{W}(%
\overset{n}{x_{j}})$. Combining this upper bound on $S$ with the lower bound
one paragraph earlier completes the proof. $\blacksquare $\smallskip

\textbf{Lemma 4.1 }\textit{If }$\mathcal{W}$ \textit{is} \textit{strictly
semi-modular} \textit{then a tight guarantee }$g$\textit{\ on the other side
of the Unanimity one touches its graph in at most one type: the equation }$%
g(x_{i})=una(x_{i})$\textit{\ has at most one solution }$x_{i}$\textit{.}

\textbf{Proof }in Section 9.3.\smallskip

We show in Section 7 that in a two person strictly super- (resp sub-)
modular problem every tight lower (resp upper) guarantee touches the
Unanimity graph. But in a semi-modular problem we know that many tight
guarantees never touch the Unanimity graph: Proposition 5.1 and Theorem 6.1
give many examples.

\subsection{Two canonical Stand Alone guarantees}

On the other side of Unanimity we find the two Stand Alone guarantees\textit{%
\ }(Definition 3.4) introduced in Section 1:%
\begin{equation}
g_{L}(x_{i})=\mathcal{W}(x_{i};\overset{n-1}{L})-\frac{n-1}{n}\mathcal{W}(%
\overset{n}{L})\text{ and }g_{H}(x_{i})=\mathcal{W}(x_{i};\overset{n-1}{H})-%
\frac{n-1}{n}\mathcal{W}(\overset{n}{H})  \label{37}
\end{equation}

In Example 2.1 $g_{L}^{+}$, $g_{H}^{+}$ are the end points of the segment $%
\mathcal{G}^{+}$. This generalises to all semi-modular problems.\smallskip

\textbf{Theorem 4.1 }\textit{Fix }$\mathcal{W}$.\textit{\ supermodular on }$%
\mathcal{X}^{n}$\textit{.}

\noindent $i)$\textit{\ The functions }$g_{L}$\textit{\ and }$g_{H}$\textit{%
\ are two tight lower guarantees of }$\mathcal{W}$\textit{.}$\smallskip $

\noindent $ii)$ \textit{The property }$g_{L}(L)=una(L)$\textit{\
characterises }$g_{L}$\textit{\ in} $\mathcal{G}^{-}$\textit{; so does }$%
g_{H}(H)=una(H)$\textit{\ for }$g_{H}$.

\noindent $iii)$ \textit{Each }$g^{-}\in \mathcal{G}^{-}$ \textit{grows
slower than }$g_{H}$\textit{\ and faster than }$g_{L}$: for any $%
x_{i}<x_{i}^{\ast }$: $g_{L}(x_{i}^{\ast })-g_{L}(x_{i})\leq
g^{-}(x_{i}^{\ast })-g^{-}(x_{i})\leq g_{H}(x_{i}^{\ast })-g_{H}(x_{i})$.$%
\smallskip $

\noindent $iv)$ \textit{For each }$g^{-}\in \mathcal{G}^{-}\diagdown
\{g_{L},g_{H}\}$\textit{\ we have}%
\begin{equation*}
g_{H}(L)<g^{-}(L)<g_{L}(L)=una(L)
\end{equation*}%
\begin{equation*}
g_{L}(H)<g^{-}(H)<g_{H}(H)=una(H)
\end{equation*}

\textit{If }$\mathcal{W}$\textit{\ is} \textit{submodular} replace lower by
upper in statement $i)$; statement $ii)$ holds unchanged; exchange $g_{H}$
and $g_{L}$ in statement $iii)$; and switch the sign of the four
inequalities in statement $iv)$.

\textbf{Proof} in Section 9.4\textbf{.\smallskip }

In the comments below we assume that $\mathcal{W}$ is supermodular, and omit
the obvious adjustment if it is submodular. For concreteness the reader can
check them in any of the Examples starting with Example 2.1.

Statement $iii)$ says that $g_{L}$ discriminates less between types, and $%
g_{H}$ discriminates more, than any other tight lower guarantee. In
particular the spread $\Delta (g)=g(H)-g(L)$\ is smallest for $g_{L}$ in $%
\mathcal{G}^{-}$and largest at $g_{H}$.

We also see that any other guarantee\textit{\ }$g^{-}\in \mathcal{G}^{-}$
crosses $g_{L}$ and $g_{H}$ only once (or in some interval).

Clearly if $\mathcal{W}(x)$ is a \textit{benefit}, $g_{L}$ favors the types $%
x_{i}$ close to $L$ who get a share close to their best case $una(x_{i})$,
and $g_{H}$ favors those close to $H$. The comparison switches if $\mathcal{W%
}(x)$ is a cost.

\subsection{Implementing the Stand Alone guarantees}

We adapt the two serial sharing rules originally introduced for the commons
problem (\cite{MoSh1}, \cite{deF}).\smallskip

\textbf{Definition 4.2}\textit{\ The increasing Serial sharing rule (Ser}$%
\uparrow $\textit{) }$\varphi ^{ser\uparrow }$\textit{\ is defined by the
combination of two properties a) it is symmetric in its variables and b) for
all }$i,j$ \textit{agent }$i$'s \textit{share does not change if }$x_{j}$%
\textit{\ changes to }$x_{j}^{\prime }$\textit{, both weakly larger than }$%
x_{i}$\textit{.}

\textit{When the agents are labelled by increasing types as }$x_{1}\leq
x_{2}\leq \cdots \leq x_{n}$\textit{\ this defines the following shares:}%
\begin{equation}
\varphi _{i}^{ser\uparrow }(x)=\frac{\mathcal{W}(x_{1},\cdots ,x_{i-1},%
\overset{n-i+1}{x_{i}})}{n-i+1}-\sum_{j=1}^{i-1}\frac{\mathcal{W}%
(x_{1},\cdots ,x_{j-1},\overset{n-j+1}{x_{j}})}{(n-j+1)(n-j)}  \label{49}
\end{equation}%
The straightforward computation is omitted; see equation (6) in \cite{MoSh1}.

The decreasing Serial rule Ser$\downarrow $ is defined by properties a) and
b)*: agent $i$' share is independent of other agents' smaller shares. It is
also given by (\ref{49}) if we label the agents by decreasing types.

In Example 2.1 the Ser$\uparrow $ rule divides the cost $x_{i}-x_{i-1}$
equally between the agents $j=i,\cdots ,n$; it is interpreted in \cite{LO}
as the Shapley value of cooperative game $v(S)=\max_{i\in S}\{x_{i}\}$ for $%
S\subseteq N$.\smallskip

\textbf{Proposition 4.2 }\textit{If }$\mathcal{W}$ is \textit{supermodular} 
\textit{the Ser}$\uparrow $\textit{\ rule implements the pair of guarantees }%
$(g_{L},una)$\textit{. The Ser}$\downarrow $\textit{\ rule implements }$%
(g_{H},una)$\textit{.}

\textit{If }$\mathcal{W}$ is\textit{\ submodular Ser}$\uparrow $ \textit{%
implements }$(una,g_{L})$\textit{.and Ser}$\downarrow $\textit{\ implements }%
$(una,g_{H})$\textit{.}

\textbf{Proof }in Section 9.5.\smallskip

Consider the cooperative game $v_{L}(S)=\mathcal{W}(\overset{S}{x},\overset{%
n-|S|}{L})$ for $S\subseteq N$. It is easy to check that its Shapley value
defines a sharing rule implementing $g_{L}$ implementing $g_{L}$, whether $%
\mathcal{W}$ is super or submodular (and a similar statement for $g_{H}$).
It does not however implement the Unanimity share.

\section{Substitutable inputs}

For a subset $S$ of $N$ we use the notation $x_{S}=\sum_{i\in S}x_{i}$.

\textbf{Definition 5.1 }\textit{A commons with substitute types (ST) is a
commons (Definition 3.0) where} $\mathcal{W}(x)=F(x_{N})$ for some
continuous function $F$ defined on $[nL,nH]$.

\textit{It is super, resp submodular if and only if }$F$\textit{\ is convex,
resp concave).\smallskip }

The Unanimity share $una(x_{i})=\frac{1}{n}F(nx_{i})$ delivers a familiar
normative critique of the venerable Average Return (AR) sharing rule $y_{i}=%
\frac{x_{i}}{x_{N}}F(x_{N})=x_{i}AR(x_{N})$ (\cite{She}, \cite{MoSh}). Say
that $\mathcal{W}$\ distributes a desirable output, $F$ is concave and $%
F(0)=0$. Intuitively the Unanimity share delivers to each agent their fair
share of the best (i. e, first) marginal returns of the technology $F$. But
agent $i$'s guarantee under AR is much smaller than $%
una(x_{i})=x_{i}AR(nx_{i})$ if $x_{i}$ is small relative to other types.%
\footnote{%
For instance if the $n-1$ types other than $x_{1}$ are $x_{j}=2x_{1}$ then $%
y_{1}=x_{1}AR((2n-1)x_{1})$, about the average return at twice the benchmark
level $nx_{1}$.}.

On the other side of (\ref{3}), the AR rule implement the SA upper guarantee
anchored at $0$: $g_{0}(x_{i})=C(x_{i})$ (because the function $AR$
decreases.), and so do the Shapley value and the serial sharing rules
(Section 4.3 just above).

To this example we apply now statement $iv)$ in Theorem 4.1. For any other
tight upper guarantee $g^{+}\in \mathcal{G}^{+}$ we have $%
g^{+}(0)>g_{0}(0)=0 $ and $g^{+}(H)<g_{0}(H)=C(H)$: a null demand agent 
\textit{may }end up paying a share of the cost generated by the other
agents! Indeed at the profile $(\overset{n-1}{0},H)$ the maximal cost share $%
g^{+}(H)$ does not cover the~total cost, so the slack must be covered by the
inactive agents.

This normative position, common to all but one tight upper guarantee,
penalises agents who choose not to consume and in doing so fail to help the
rest of the agents. This make no sense if they are pooling demands to
benefit from a discounted price, but does make sense for the consumption of
vaccination and other insurances with positive externalities on my fellow
agents. If we wish to maximise this pressure to consume, the best cap on
cost shares is $g_{H}$, again by $iv)$ in Theorem 4.1: $g_{H}(L)$ is largest
in $\mathcal{G}^{+}$ and $g_{H}(H)$ is smallest.

Congestion effects in queues, as in \cite{She}, or polluting activities
generate a convex cost function $C$ and negative consumption externalities.
The symmetric impact of all but one tight lower guarantee is to reward the
abstemious agents: for a small enough consumption their net cash transfer
may be positive.See the numerical Example 5.1 below.

\subsection{The $n$ canonical simple guarantees}

Back to the general Definition 5.1 we do not assume that $F$\ is monotonic,
nor that $0$ plays a special role in $\mathcal{X}$.

The first set of tight guarantees connecting the Stand Alone $g_{L}$ and $%
g_{H}$ is a sequence of $n-2$ simple guarantees (Definition 3.3)~illustrated
in the three Examples of Section 5.3.$\smallskip $

\textbf{Proposition 5.1 }\textit{Suppose}\textbf{\ }$n\geq 3$ \textit{and
fix an arbitrary ST\ commons }$([L,H],F)$. \textit{Consider the }$n$ \textit{%
functions }%
\begin{equation*}
g_{\ell ,h}(x_{i})=F(x_{i}+\ell L+hH){\large -}\frac{1}{n}[\ell F((\ell
+1)L+hH)+hF(\ell L+(h+1)H)]
\end{equation*}
where\textit{\ }$\ell ,h$\textit{\ are integers s. t. }$0\leq \ell ,h\leq
n-1 $\textit{\ and }$\ell +h=n-1$\textit{. Note that }$g_{n-1,0}=g_{L}$%
\textit{\ and }$g_{0,n-1}=g_{H}$\textit{.}$\smallskip $

\textit{\noindent }$i)$\textit{\ If }$F$\textit{\ is convex (resp concave) }$%
g_{\ell ,h}$\textit{\ is a simple tight lower (resp upper) guarantee of }$F$%
\textit{. The profile }$(x_{i},\overset{\ell }{L},\overset{h}{H})$\textit{\
is a contact profile of }$g_{\ell ,h}$\textit{\ for all }$x_{i}$\textit{.}$%
\smallskip $

\textit{\noindent }$ii)$\textit{\ If }$F$\textit{\ is strictly convex (resp
strictly concave) the }$n-2$\textit{\ guarantees other than }$g_{L}$\textit{%
\ and }$g_{H}$\textit{\ do not touch the Unanimity one (their graphs do not
intersect). The gap }$|una(x_{i})-g_{\ell ,h}(x_{i})|$\textit{\ is minimal
at the type }$x_{i}=\frac{\ell }{n-1}L+\frac{h}{n-1}H$\textit{.}$\smallskip $

To check that $g_{\ell ,h}$ takes the form (\ref{22}) use $c=(\overset{\ell }%
{L},\overset{h}{H})$.

\textbf{Proof }in Section 9.6\textbf{\smallskip }

By statement $ii)$ the choice of the parameters $\ell ,h$ can be guided by
the choice of the benchmark type of which the share depends the least on
other types. The three examples.

There is perhaps a sequence of natural division rules \textquotedblleft
between\textquotedblright\ Ser$\uparrow $ and Ser$\downarrow $ to implement
the $n-2$ lower guarantees between $g_{L}$ and $g_{H}$. Short of discovering
one we can use the moving average rules described after Lemma 3.6.

\subsection{Tangent and hybrid guarantees}

If the general function $\mathcal{W}$ (where the inputs may not be
substitute) is globally convex\footnote{%
Recall that convexity and super or sub modularity are not logically related.}
and differentiable in $[L,H]^{N}$ the tangent $\theta _{a}$ at a point $%
(a,una(a))$ to the graph of $una$ is a lower guarantee of $\mathcal{W}$.
Indeed $\theta _{a}(x_{i})=\frac{1}{n}\mathcal{W}(\overset{n}{a})+\partial
_{1}\mathcal{W}(\overset{n}{a})(x_{i}-a)$ so the LH inequalities in (\ref{3}%
) for $\theta _{a}$ are: $\sum_{N}\theta _{a}(x_{i})=\mathcal{W}(\overset{n}{%
a})+\partial _{1}\mathcal{W}(\overset{n}{a})(x_{N}-na))\leq \mathcal{W}(x)$
for all $x$, precisely the tangent hyperplane inequality of $\mathcal{W}$ at 
$(\overset{n}{a})$ because $\mathcal{W}$ is symmetric.

If $\mathcal{W}(x)=F(x_{N})$ with $F$ convex we find that a tangent to the
graph of $una$ is in fact a \textit{tight} lower guarantees \textbf{if} the
tangency point is in a certain interior subinterval of $[L,H]$ (described
shortly).

For the formal statement recall that if $F$ is either convex or concave in $%
[L,H]$, it has well defined left and right derivatives. We write $\frac{dF}{%
dx}(z)$ the closed interval they define (almost everywhere a single point)
and $\Theta _{a}$ for the set of tangent(s) to the Unanimity graph at $na$
when the slope varies in $\frac{dF}{dx}(na)$.\smallskip

\textbf{Proposition 5.2}: \textit{If }$F$ \textit{is convex in }$[nL,nH]$ 
\textit{the supermodular commons }$\mathcal{W}(x)=F(x_{N})$\textit{\ admits
the following tight lower guarantees }$g_{a}$\textit{\ where }$a\in \lbrack
L,H]$\textit{; they connect }$g_{L}$ to $g_{H}$.$\smallskip $

Case 1:\textit{\ If }$\frac{n-1}{n}L+\frac{1}{n}H\leq a\leq \frac{1}{n}L+%
\frac{n-1}{n}H$\textit{\ any }$g_{a}=\theta _{a}\in \Theta _{a}$\textit{.}$%
\smallskip $

Case 2: \textit{If} $L\leq a\leq \frac{n-1}{n}L+\frac{1}{n}H$\textit{\ then }%
$g_{a}=\theta _{a}\in \Theta _{a}$ \textit{in }$[L,na-(n-1)L]$ and $%
g_{a}=g_{L}+(n-1)(una(L)-\theta _{a}(L))$ in $[na-(n-1)L,H]$.

Case 3:\textit{\ If }$\frac{1}{n}L+\frac{n-1}{n}H\leq a\leq H$ \textit{.Then 
}$g_{a}=g_{H}+(n-1)(una(H)-\theta _{a}(H))$ \textit{in} $[L,na-(n-1)H]$%
\textit{\ and } $g_{a}=\theta _{a}\in \Theta _{a}\mathit{\ }$in $%
[na-(n-1)H,H]$.

\textit{If }$F$\textit{\ is concave in }$[nL,nH]$\textit{\ the same
definitions produce tight upper guarantees of }$\mathcal{W}$.

\textbf{Proof }in Section 9.7.\textbf{\smallskip }

For $n\geq 3$ all the tangents touching the graph of $una$ along the
symmetric subinterval of $[L,H]$ of size $\frac{n-2}{n}$ (Case 1) are tight
lower guarantees of $F$. For $n=2$ Case 1 is an endpoint of Cases 2 and 3.

In Cases 2,3 the guarantees concatenate a tangent to $una$ with the
translated of one of the two SA guarantees.$\smallskip $

Here is an important consequence of Theorem 7.1 below for a semi-modular $%
\mathcal{W}$. At a given contact point on the graph of its Unanimity
function we expect many other (a large infinity) of tight guarantees. So the
tight guarantees in Propositions 5.1, 5.2 far from exhaust the set $\mathcal{%
G}^{-}$.\smallskip

\subsection{Three examples}

\textbf{Example 5.1} \textit{Sharing the cost of a public bad }

Each agent engages in a potentially polluting activity at a level $x_{i}$ in 
$\mathcal{X}=[0,2d]$. The convex cleaning cost is $F(x_{N})=(x_{N}-nd)_{+}$:
total pollution $x_{N}$ is harmless up to $nd$, beyond which it must be
cleaned at cost $1$.

The Unanimity cost share $una(x_{i})=(x_{i}-d)_{+}$ is the single tight cap
on cost shares for the different types. A \textquotedblleft
clean\textquotedblright\ type, $x_{i}\leq d$, will never pay but could be
paid; the cost, if any, must be shared by the \textquotedblleft
dirty\textquotedblright\ types $x_{j}>d$; each such type may pay up to the
full cost of its pollution in excess of the benchmark $d$.

The Average Cost rule $\varphi _{i}^{AC}(x)=\frac{x_{i}}{x_{N}}%
(x_{N}-nd)_{+} $ violates the Unanimity cap and is far from tight: it
charges a positive type, even a very clean one, as soon as total pollution
is costly; this charge can be as high as about $\frac{1}{2}x_{i}$ in the
worst case.\footnote{%
The same critique of AC applies to the Shapley value of the cooperative game 
$v(T)=(x_{T}-nd)_{+}$ for $T\subseteq N$.}

The SA guarantee $g_{L}$ (minimal cost share) is identically zero: nobody
pays anything if $x_{N}\leq nd$ and no type is ever paid for not polluting.
The SA guarantee $g_{H}$ is, as expected, quite different: $%
g_{H}(x_{i})=x_{i}-d$, so a dirty agent $i$ pays his worst (Unanimity) cost 
\textit{for sure} ($g_{H}(x_{i})=una(x_{i})$in $[d,2d]$). If total pollution
incurs no cost, $x_{N}\leq nd$, the tax $x_{i}-d$ goes to the clean agents;
a zero polluter can receive as much as $d$ in cash. Under $g_{H}$ a clean
agent $j$ is effectively selling her unused pollution credit $d-x_{j}$ to
absorb\ the pollution of the dirty ones.

If $n=2m+1$ is odd, the $n-2$ simple lower guarantees in Proposition 5.1
reduce to a single one $g_{m,m}$: $g_{m,m}(x_{i})=(x_{i}-d)_{+}-\frac{m}{n}d$%
.\footnote{%
Check that $g_{\ell ,h}=g_{L}$ if $h<\ell $, $=g_{H}$ if $h>\ell $.} It
strikes a reasonable compromise between $g_{L}$ and $g_{H}$: an agent is
taxed for sure only if her pollution is larger than $(1+\frac{m}{n})d\simeq 
\frac{3}{2}d$; the dirtiest type $2d$ is taxed at least $\frac{m+1}{n}d$,
and the zero polluter is paid at most $\frac{m}{n}d$.

Proposition 5.2 delivers a continuum of tight lower guarantees. It is the
entire cone of tangents to the Unanimity graph, all touching it at its kink $%
(d,0)$: $g_{\lambda }(x_{i})=\lambda (x_{i}-d)$ for $0\leq \lambda \leq 1$;
we get $g_{L}$ for $\lambda =0$ and $g_{H}$ for $\lambda =1$. Pollution in
excess of $d$ is taxed (at least) at the fixed rate $\lambda $ and the same
rate applies to the maximal subsidy of clean types.

\textbf{Figure 2} illustrates these guarantees for $n=3$ and $d=1$. The
simple $g_{1,1}(x_{i})=(x_{i}-1)_{+}-\frac{1}{3}$ is a vertical translation
of the Unanimity graph; the tangent guarantees are all the lines through $%
(1,0)$ with slope in $[0,1]$.\smallskip

\textbf{Example 5.2} \textit{Complementary inputs}

The four agents join inputs to complete a project that returns the product
of these inputs: $\mathcal{W}(x)=x_{1}x_{2}x_{3}x_{4}$ where $\mathcal{X}%
=[1,5]$. In this supermodular production function the mutual positive
externalities are massive: a relative shift of $t\%$ to the input of any
agent has the same impact on total output.

We have $una(x_{i})=\frac{1}{4}x^{4}$. The contrast between the two lower
guarantees $g_{L}(x_{i})=x_{i}-\frac{3}{4}$ and $%
g_{H}(x_{i})=125(x_{i}-3.75) $ is now enormous. The former, $g_{L}$,
guarantees a positive share of output even to the smallest input, and only
the share $g_{L}(5)=4.25$ to the largest input $x_{i}=5$. Contrast this with 
$g_{H}(x_{i})=125(x_{i}-3.75)$, guaranteeing the share $%
g_{H}(5)=una(5)=156.25$ to the input $5$, exactly $\frac{1}{4}$ of the
output $\mathcal{W}(\overset{4}{5})$; then at the profile $x=(1,\overset{3}{5%
})$ the lazy type $1$ must pay a tax of $343.75$!

To find reasonable compromises between these two extremes, the change of
variable $x_{i}=e^{z_{i}}$ turns the function $\mathcal{W}$ to $\mathcal{W}%
^{\blacklozenge }(z)=e^{z_{N}}$ to which we apply Propositions 5.1. The two
guarantees $g_{1,2}^{\blacklozenge }$ and $g_{2,1}^{\blacklozenge }$ of $%
\mathcal{W}^{\blacklozenge }$ become $g_{1,2}(x_{i})=25(x_{i}-2.75)$ and $%
g_{2,1}(x_{i})=5(x_{i}-1.75)$ for $\mathcal{W}$. The worst punitive tax
under $g_{1,2}$ is only $43.25$ and fewer types can ever be taxed. Under $%
g_{2,1}$ this tax is at most $3.75$

More compromises come from Prop 5.2. Because the change of variables $%
x_{i}=e^{z_{i}}$ is smooth, the tangent guarantees $g_{\alpha
}^{\blacklozenge }$ of $\mathcal{W}^{\blacklozenge }$ (Case 1) turn into
tangents $g_{a}$ to the Unanimity graph of $\mathcal{W}$: $%
g_{a}(x_{i})=a^{4}(\frac{1}{4}+\ln (\frac{x_{i}}{a}))$ for $5^{\frac{1}{4}%
}\leq a\leq 5^{\frac{3}{4}}$.\footnote{%
coming from $g_{\alpha }^{\blacklozenge }(z_{i})=e^{4\alpha }(\frac{1}{4}%
+z_{i}-\alpha )$, for $\frac{1}{4}\ln (5)\leq \alpha \leq \frac{3}{4}\ln (5)$%
.
\par
{}} For instance if we choose $a=5^{\frac{1}{2}}e^{\frac{1}{4}}\simeq 2.87$
we have $g_{a}(5)=|g_{a}(1)|\simeq 54.7$: the worst tax to the laziest type
equals the guaranteed share of the most active one.\smallskip

\textbf{Example 5.3} \textit{Sharing the cost of the variance}

The agents choose a type $x_{i}$ in $[0,1]$ and must share ($n$ times) the
variance of their distribution: $\mathcal{W}(x)=\sum_{N}x_{i}^{2}-\frac{1}{n}%
{\large (}\sum_{N}x_{i}{\large )}^{2}$.We can rewrite the submodular
function $\mathcal{W}$ as $\sum_{N}(x_{i}-\frac{1}{n}x_{N})^{2}$ and
interpret it as the total quadratic transportation cost of the types to
their mean. This suggests the natural cost sharing rule $\varphi _{i}^{\ast
}(x)=(x_{i}-\frac{1}{n}x_{N})^{2}$. It satisfies the reasonable Unanimity
share $una(x_{i})\equiv 0$: no type is ever subsidised by the others.
However, we see below that the upper guarantee of $\varphi ^{\ast }$ is 
\textit{very} far from tight.

By statement $i)$ of Lemma 3.4, the tight upper guarantees of $\mathcal{W}$
obtain from those of $\mathcal{W}^{\blacklozenge }(x)=-{\large (}%
\sum_{N}x_{i}{\large )}^{2}$ by simply adding $x_{i}^{2}$.

A first set of guarantees in $\mathcal{G}^{+}$ pick a benchmark location $a$
that is free, and charge the other types for the travel cost to $a$. The two
canonical SA are $g_{L}(x_{i})=\frac{n-1}{n}x_{i}^{2}$ and $g_{H}(x_{i})=%
\frac{n-1}{n}(x_{i}-1)^{2}$ (here the travel cost is dicounted). Then the
tangents in Case 1 of Proposition 5.2 (applied to $\mathcal{W}%
^{\blacklozenge }$) give $g_{a}^{+}(x_{i})=(x_{i}-a)^{2}$ for $\frac{1}{n}%
\leq a\leq \frac{n-1}{n}$.

The $n-2$ upper guarantees other than $g_{L},g_{H}$ (Proposition 5.1)
indexed by $h\in \lbrack n-2]$ are $\overline{g}_{h}^{+}(x_{i})=\frac{n-1}{n}%
(x_{i}-\frac{h}{n-1})^{2}+\delta _{h}$ where $\delta _{h}=\frac{h(n-1-h)}{%
n^{2}(n-1)}$. The worst cost share is positive for all types but small near $%
\frac{h}{n-1}$ as $\delta _{h}\leq \frac{1}{4n}$ for all $h$. Note the
similarity between the guarantee $g_{a}^{+}$ and $\overline{g}_{h}^{+}$ for $%
a=\frac{h}{n-1}$: $g_{h}^{+}$ is $\frac{n-1}{n}$ times flatter than $%
g_{a}^{+}$ but unlike $g_{a}^{+}$ it never vanishes.

Returning to the sharing rule $\varphi ^{\ast }$ we compute its upper
guarantee $g^{+}(x_{i})=\frac{n-1}{n}(x_{i}-1)^{2}$ if $x_{i}\leq \frac{1}{2}
$, $=\frac{n-1}{n}x_{i}^{2}$ if $x_{i}\geq \frac{1}{2}$. This is the maximum
of $g_{L}$ and $g_{H}$, and much larger everywhere than the symmetric
tangent guarantee $g_{\frac{1}{2}}(x_{i})=(x_{i}-\frac{1}{2})^{2}$.

\section{Rank-separable functions}

The \textit{decreasing order statistics} of the profile $x\in \lbrack
L,H]^{N}$ is written $(x^{k})_{k=1}^{n}$ with $x^{1}=\max_{i}\{x_{i}\}$ and $%
x^{n}=\min_{i}\{x_{i}\}$. The statement \textquotedblleft $x_{i}$ is of rank 
$k$ in profile $x$\textquotedblright\ means that $x_{i}$ appears at rank $k$
for some weakly decreasing ordering of the coordinates of $x$.\smallskip

\textbf{Definition 6.1} \textit{The function} $\mathcal{W}$ \textit{on} $%
[L,H]^{N}$ \textit{is rank-separable if it takes the form }$\mathcal{W}%
(x)=\sum_{k=1}^{n}w_{k}(x^{k})$\textit{\ for some continuous }functions $%
w_{k}$\textit{\ on }$[L,H]$\textit{, }$k\in \lbrack n]$.\smallskip

\textbf{Lemma 6.1 }\textit{The rank-separable function} $\mathcal{W}$ 
\textit{is supermodular if and and only if }$w_{k}$\textit{\ grows weakly
slower than }$w_{k+1}$\textit{\ in }$[L,H]$: $w_{k}(y)-w_{k}(z)\leq
w_{k+1}(y)-w_{k+1}(z)$ \textit{for all }$z\leq y$\textit{\ and }$k\in
\lbrack n-1]$.

\textit{It is submodular if and only if }$w_{k}$\textit{\ grows weakly
faster than }$w_{k+1}$.

\textbf{Proof }in Section 9.8.\smallskip

If the function $\mathcal{W}$ is additive in the ranked types as $\mathcal{W}%
(x)=\sum_{k=1}^{n}\lambda _{k}x^{k}$, it is supermodular (resp submodular)
if and only if the sequence $(\lambda _{k})$ increases (resp decreases)
weakly. So the $k$-rank function $\mathcal{W}(x)=x^{k}$ with $2\leq k\leq
n-1 $ is neither.

The rank-separable and semi-modular functions are the largest class for
which we know precisely all the tight guarantees opposite to the Unanimity:%
\textit{\ }they are the simple guarantees $g_{c}$ in Definition 3.3.
Conversely the only semi-modular functions of which all tight guarantees are
simple, are the rank-separable ones.\smallskip

\textbf{Theorem 6.1 }\textit{Fix a semi-modular function }$\mathcal{W}$. 
\textit{The two following statements are equivalent:}

\noindent $i)$ \textit{The function} $\mathcal{W}$ \textit{is rank-separable.%
}

\noindent $ii)$ O\textit{n the other side of the Unanimity guarantee, for
all }$c\in \lbrack L,H]^{n-1}$ \textit{the function }$g_{c}$\textit{\ (\ref%
{22}) is a (tight) simple guarantee of }$\mathcal{W}$.

\textit{Then the function }$\mathcal{W}$\textit{\ has no other tight
guarantee}.

The long\textbf{\ Proof }is\ in Section 9.9\textbf{.}\smallskip

For this class of functions the set of guarantees opposite to $una$ is of
dimension at most $n-1$. It is exactly $n-1$ if the inequalities in Lemma
6.1 are strict, but not for $\mathcal{W}(x)=\max_{i}\{x_{i}\}$ where $g_{c}$
only depends on $\max_{j}\{c_{j}\}$ (Lemma 2.1).

\subsection{Three examples}

The first two functions $\mathcal{W}$ below are additive with respect to the
ordered types $x^{k}$. By Lemma 6.1 and the definition of $g_{c}$ this
implies that every tight guarantee opposite to $una$ is piecewise linear and
concave, resp convex if $\mathcal{W}$ is super, resp submodular.\smallskip

\textbf{Example 6.1 }\textit{Team work and Queuing}

From the comment just after Lemma 6.1, the function $\mathcal{W}%
(x)=x^{1}+2x^{2}+\cdots +nx^{n}$ is supermodular on $[0,H]^{n}$. We give two
interpretations.

\textbf{Story 1: }\textit{Team work with increasing returns}

When $k$ agents work as a team their productivity is $\frac{1}{2}k(k+1)$ per
hour. Agent $i$ can only work for $x_{i}$ hours. To maximise the total
output we let the full team working from $x^{n}$ hours, then all but $n$ for 
$x^{n-1}-x^{n}$ hours, etc..., resulting in\ the output above. How should we
divide it?

\textbf{Story 2}: \textit{Queuing with cash compensation }(\textbf{\cite{Man}%
, \cite{Chu} })

A single server processes $n$ jobs, one per agent, each job takes one day,
and agent $i$'s waiting cost is $x_{i}$ per day. Serving high cost agents
first gets the minimal total cost $\mathcal{W}(x)$. What cash compensations
from the impatient agents to the patient ones are fair?

The Unanimity upper guarantee is $una(x_{i})=\frac{1}{2}(n+1)x_{i}$ and the
first canonical SA guarantee is $g_{0}(x_{i})=x_{i}$. Under the tight pair $%
(g_{0},una)$ a positive work time guarantees a positive share of output in
story 1; in story 2 even the net cost (waiting cost plus cash transfer) of
even a very patient agent is positive. The latter accords with the axiomatic
literature on the queuing model. For instance the elegant sharing rule
proposed in \textbf{\cite{Man}} implements the pair $(g_{L},una)$: it makes $%
i$ pay, in addition to $x_{i}$, one half of $x_{j}$ for each agent $j$ that $%
i$ displaces in the queue.

Any other tight lower guarantee has $g^{-}(0)<0$: an absentee in story 1 may
be taxed and a very patient one in story 2 can be compensated more than his
waiting cost. We can adjust the threshold at which $g^{-}(x_{i})$ changes
sign by choosing the contact profile $(\overset{n-1}{c_{0}})$ of a Stand
Alone lower guarantee (Definition 3.4): $g_{c_{0}}^{-}(x_{i})=\min \{nx_{i}-%
\frac{n-1}{2}c_{0},x_{i}+\frac{n-1}{2}c_{0}\}$. This is the simplest segment
of guarantees linking $g_{0}$ and $g_{H}$. Their critical threshold can be
anywhere in $[0,\frac{n-1}{2n}H]$ but not higher.\smallskip

\textbf{Example 6.2 }\textit{Sharing the cost of the spread}

The types $x_{i}$ vary in the interval $[0,1]$\ and the agents must share
the submodular cost of the spread $\mathcal{W}(x)=x^{1}-x^{n}$. For instance
types represent a location and $\mathcal{W}(x)$ is the cost of a road
connecting the agents. Or type $x_{i}$ is the date at which $i$ shows up to
pick a parcel, and $\mathcal{W}(x)$ is (a proxy for) the cost of keeping the
storage facility open.

This problem is similar to Example 5.3 where the cost is the variance of the
types. But here we find \textit{all} the tight guarantees of the function $%
\mathcal{W}$, a far cry from the situation in that Example.

Unamnimity share is $una(x_{i})\equiv 0$ as in Exampe 5.3: everyone's best
case is to pay nothing, but nobody can make a profit. The SA upper guarantee
(Definition 3.4) with parameter $(\overset{n-1}{c_{0}})$ is $%
g_{c_{0}}^{+}(x_{i})=|x_{i}-c_{0}|$: type $c_{0}$ is free, and a single
agent deviating from $c_{0}$ pays the full connection cost to this
benchmark. For $n=2$ there are no other tight guarantees.

For $n\geq 3$ and any $c\in \lbrack 0,1]^{n-1}$ we compute easily the simple
guarantee $g_{c}^{+}$ (Definition 3.3): it depends only upon the largest and
smallest parameters $c_{\max }$ and $c_{\min }$: $g_{c}^{+}(x_{i})=\frac{1}{n%
}(c_{\max }-c_{\min })+\max \{0,x_{i}-c_{\max },c_{\min }-x_{i}\}$. All
types in the safe interval $[c_{\min },c_{\max }]$ pay at most their fair
share of its length, other types may add to this the full cost of connecting
their type to the safe interval. If $[c_{\min },c_{\max }]=[0,1]$ then $%
g_{c}^{+}$ is $\frac{1}{n}$ for all types, implemented by the equal split
rule $\varphi _{i}(x)=\frac{1}{n}(x^{1}-x^{n})$. To implement a general $%
g_{c}^{+}$ we can for instance share the cost $x^{1}-c_{\max }$, if any,
equally between the types above $c_{\max }$, do the same for the types below 
$c_{\min }$ and share the remaining cost equally among all types.\smallskip

\textbf{Remark 6.1} We interpreted Example 5.3 as the division of quadratic
total travel costs to an optimally located facility. With linear travel cost
this become the cost of traveling to the median $x^{m+1}$ of the types
(where $n=2m+1$), computed as $\mathcal{W}(x)=\sum_{k=1}^{m}x^{k}-\sum_{\ell
=m+2}^{n}x^{\ell }$, another rank-separable function.\smallskip

\textbf{Example 6.3}\textit{\ Production with quota}

Fix an integer quota $q$ s.t. $2\leq q\leq n-1$. Agent $i$ inputs the effort 
$x_{i}$ and to achieve the output level $F(e)$ we need at least $q$ agents
contributing an effort at least $e$: $\mathcal{W}_{q}(x)=F(x^{q})$. Assume
that $F$ is continuous and strictly increasing. By Lemma 6.1 $\mathcal{W}%
_{q} $ is \textbf{not} semi-modular.

The Unanimity $una(x_{i})=\frac{1}{n}F(x_{i})$ is neither a lower guarantee
nor an upper guarantee. There is precisely a one dimensional set of tight
guarantees on each side of (\ref{3}), parametrised respectively by $p^{-}$
and $p^{+}$ varying in $[L,H]$. For $\mathcal{G}_{q}^{+}$ they are: $%
g_{p^{+}}^{+}(x_{i})=\frac{1}{n}F(p^{+})+\frac{1}{q}(F(x_{i})-F(p^{+}))_{+}$
and for $\mathcal{G}_{q}^{-}$: $g_{p^{-}}^{-}(x_{i})=\frac{1}{n}F(p^{-})+%
\frac{1}{n-q+1}(F(x_{i})-F(p^{-}))$. The proof, omitted for brevity, mimicks
that of Proposition 2.1.

If $p^{+}=p^{-}=\widetilde{p}$ this benchmark\ level of effort guarantees
the share $\frac{1}{n}F(\widetilde{p})$. If the actual input $x^{q}$ is
below $\widetilde{p}$ the hard working types at or above $\widetilde{p}$ get
at least $\frac{1}{n}F(\widetilde{p})$, therefore the slacker at or below $%
x^{q}$\ get on average less than $\frac{1}{n}F(\widetilde{p})$.

\section{Two person strictly semi-modular problems}

The key to our characterisation result is a precise description, in the next
two Lemmas, of the contact set of tight guarantees on the other side of $una$%
. Fixing a semi-modular function $\mathcal{W}(x_{1},x_{2})$ and a tight
guarantee $g$ on either side of (\ref{3}), its contact correspondence $%
\gamma $ maps $[L,H]$ into itself: $\gamma (x_{1})=\{x_{2}\in \lbrack
L,H]:g(x_{1})+g(x_{2})=\mathcal{W}(x_{1},x_{2})\}$ for all $x_{1}\in \lbrack
L,H]\}$. It is non empty (Lemma 3.2) and we write its graph $\Gamma (\gamma
) $.\smallskip

\textbf{Lemma 7.1 }\textit{Fix }$\mathcal{W}$ \textit{supermodular,} $g\in 
\mathcal{G}^{-}$ \textit{and suppose that }$\Gamma (\gamma )$\textit{\
contains }$(x_{1},x_{2})$\textit{\ and }$(x_{1}^{\ast },x_{2}^{\ast })$%
\textit{\ s.t. }$(x_{1},x_{2})\ll (x_{1}^{\ast },x_{2}^{\ast })$\textit{.
Then }$\Gamma (\gamma )$ \textit{contains }$(x_{1},x_{2}^{\ast
}),(x_{1}^{\ast },x_{2})$\textit{\ as well and }$\mathcal{W}$\textit{\ is
not strictly supermodular. If }$\mathcal{W}$\textit{\ is submodular replace }%
$\mathcal{G}^{-}$ \textit{by} $\mathcal{G}^{+}$.

\textbf{Proof} Sum up the two equalities defining $\gamma $ for $%
(x_{1},x_{2})$ and $(x_{1}^{\ast },x_{2}^{\ast })$, then combine the
resulting equality with the supermoduar inequality (\textit{\ref{38}) }to get%
\textit{\ }$\{g(x_{1})+g(x_{2}^{\ast })\}+\{g(x_{1}^{\ast })+g(x_{2})\}\leq 
\mathcal{W}(x_{1},x_{2}^{\ast })+\mathcal{W}(x_{1}^{\ast },x_{2})$. By (\ref%
{38}) again, this inequality is an equality and the desired contradiction
follows. In fact $\mathcal{W}$ is then separably additive inside the
rectangle $[x_{1},x_{1}^{\ast }]\times \lbrack x_{2},x_{2}^{\ast }]$. $%
\blacksquare $\smallskip

\textbf{Lemma 7.2 }\textit{Fix a strictly supermodular function }$\mathcal{W}
$\textit{\ and a tight guarantee }$g\in \mathcal{G}^{-}$ \textit{-- or a
submodular }$\mathcal{W}$\textit{\ and }$g\in \mathcal{G}^{+}$ --\textit{\
with contact correspondence }$\gamma $. \textit{Then}$\smallskip $

\noindent $i)$\textit{\ }$\Gamma (\gamma )$\textit{\ is symmetric:} $%
x_{2}\in \gamma (x_{1})\Longleftrightarrow x_{1}\in \gamma (x_{2})$ \textit{%
for all }$x_{1},x_{2}$.$\smallskip $

\noindent $ii)$ $\gamma $\textit{\ is convex valued:} $\gamma
(x_{1})=[\gamma ^{-}(x_{1}),\gamma ^{+}(x_{1})]$\textit{,\ single-valued
a.e., and upper-hemi-continuous (its graph is closed).}$\smallskip $

\noindent $iii)$\textit{\ }$\gamma ^{-}$\textit{\ and }$\gamma ^{+}$\textit{%
\ are weakly decreasing and }$x_{1}\leq x_{1}^{\ast }\Longrightarrow \gamma
^{-}(x_{1})\geq \gamma ^{+}(x_{1}^{\ast })$; $\gamma $ \textit{is the u.h.c.
closure of both }$\gamma ^{-}$ and $\gamma ^{+}$.$\smallskip $

\noindent $iv)$\textit{\ }$\gamma (L)$ \textit{contains }$H$\textit{\ and }$%
\gamma (H)$\textit{\ contains }$L$.$\smallskip $

\noindent $v)$ $\gamma $\textit{\ has a unique fixed point }$a$: $a\in
\gamma (a)$\textit{, and }$a$\textit{\ is an end-point of }$\gamma (a)$%
\textit{.\smallskip }

\textbf{Proof }in Appendix 9.10.\smallskip

Note that property $iv)$ simply means $g(L)+g(H)=\mathcal{W}(L,H)$. In fact
this property holds for \textit{all} semi-modular functions, even not
strictly so.\footnote{%
Pick a contact type $x$ for type $L$ and $y$ for $H$; then sum up the two
equations in Lemma 3.2 and rearrange them as $\mathcal{W}(L,x)+\mathcal{W}%
(y,H)=g(L)+g(H)+g(x)+g(y)\leq \mathcal{W}(L,H)+\mathcal{W}(x,y)$: comparing
with (\ref{38}) we have in fact an equality and are done.}

After picking a correspondence $\gamma $ as in Lemma 7.2, we construct the
tight guarantee of which $\gamma $ describes the contact set by integrating
the differential equation $\frac{dg}{dx_{i}}(x_{i})=\frac{\partial \mathcal{W%
}}{\partial x_{i}}(x_{i},\gamma (x_{i}))$ explained in Section 9.11.\textit{%
\smallskip }

\textbf{Theorem 7.1 }\textit{Fix a strictly super (resp. sub) modular
function }$\mathcal{W}$\textit{, continuously differentiable in }$[L,H]^{2}$%
\textit{.}

\noindent $i)$\textit{\ For any correspondence }$\gamma $\textit{\ as in
Lemma 7.2, the following equation}%
\begin{equation}
g(x_{1})=\int_{a}^{x_{1}}\partial _{1}\mathcal{W}(t,\gamma (t))dt+una(a)
\label{12}
\end{equation}%
\textit{defines a tight lower guarantee} $g$\textit{\ in} $\mathcal{G}^{-}$ 
\textit{(resp. a tight upper one in} $\mathcal{G}^{+}$\textit{).}$\smallskip 
$

\noindent $ii)$ \textit{Conversely if }$g$\textit{\ is a tight guarantee in} 
$\mathcal{G}^{-}$ \textit{(resp.} $\mathcal{G}^{+}$\textit{) with contact
correspondence }$\gamma $,\textit{\ then }$g$ \textit{takes the form (\ref%
{12}).\smallskip }

\textbf{Corollary: }\textit{If }$n=2$\textit{\ each tight guarantee on the
other side of }$una$\textit{\ touches its graph (is tangent to it if smooth)
at a unique point.}

\textbf{Proof }in Sections 9.11 and 9.12\textbf{.}\smallskip

The contact correspondences of the two SA guarantees $g_{L},g_{H}$ follow
respectively the lower and upper edges of the square $[L,H]^{2}$: e. g., $%
\gamma _{L}(0)=[0,1]$, $\gamma _{L}(]0,1])=0$. A natural compromise between
these two follows the anti-diagonal of $[L,H]^{2}$: $\gamma
(x_{1})=L+H-x_{1} $. With the notation $d=\frac{1}{2}(L+H)$ this gives $%
g_{d}(x_{i})=una(d)+\int_{d}^{x_{i}}\partial _{1}\mathcal{W}(t,2d-t)dt$. In
the commons model of Section 5 this is precisely the unique tangent tight
guarantee (Case 1 in Prop. 5.2).

By Lemma 7.2 the general construction of a correspondence $\gamma $, and its
associated tight guarantee, goes as follows. After selecting the type $a$ at
which $\gamma $ crosses the diagonal of $[L,H]^{2}$, we pick \textit{any }%
decreasing single-valued function $\overline{\gamma }$ from $[L,a]$ into $%
[a,H]$ mapping $L$ to $H$ and $a$ to itself, then fill the (countably many)
jumps down to create the correspondence $\gamma $ of which the graph
connects $(L,H)$ to $(a,a)$, and finally extend $\gamma $ to $[a,H]$ by
symmetry of its graph around the diagonal of $[L,H]^{2}$. Clearly two
different $\gamma $-s give two different guarantees.

So the sets $\mathcal{G}^{\pm }$ on the other side of Unanimity are
parametrised by a functional space and their dimension is infinite.

Contrast Theorem 7.1 with Theorem 6.1m where we saw that the dimension of
these sets $\mathcal{G}^{\pm }$ is finite for rank-separable functions. They
are indeed very far from \textit{strictly} semi-modular: in the open subsets
of $[L,H]^{N}$ where the ordering of the coordinates is strict and constant
they are additively separable: inequality (\ref{38}) is an equality in many
open sets.

\section{Some open questions}

\textbf{Q1} \textit{Implementing tight guarantees by single valued sharing
rules}

Expanding the question raised at the end of Section 5.1, what natural
sharing rules will implement the tangent guarantees in Case 1 of Proposition
5.2? Or the simple guarantees of rank-separable problems? Or a general two
person guarantee given by (\ref{12})?\smallskip

\noindent \textbf{Q2} \textit{Generalising Theorem 7.1 for} $n\geq 3$.

The key for this result is the description of the contact correspondence of
any tight guarantee in Lemma 7.2. We could not gain a similar understanding
of this correspondence with three or more agents.

Note that for $n=2$ the contact correspondence of \textit{every} tight
guarantee $g$ in $\mathcal{G}^{\varepsilon }$ intersects the diagonal: the
graph of $g$ touches that of $una$. This gives the starting point of the
integral equation (\ref{12}). But for $n\geq 3$ Proposition 5.1 and Theorem
6.1 describe many tight guarantees of which the contact set does not
intersect the diagonal.\smallskip

\noindent \textbf{Q3} \textit{Multi-dimensional types.}

All definitions and results of Section 3 are preserved if the type set $%
\mathcal{X}$ is a rectangle $[L,H]$ in $%
\mathbb{R}
^{K}$. An obstacle to develop the multidimensional analysis is the following
challenging question.

The following claim is obvious from the definitions and Lemma 3.5. Suppose
each type has two components $x_{i}=(x_{i}^{a},x_{i}^{b})\in \mathcal{X}%
_{a}\times \mathcal{X}_{b}=\mathcal{X}$ and pick two functions $\mathcal{W}%
_{a}$ on $\mathcal{X}_{a}^{[n]}$ and $\mathcal{W}_{b}$ on $\mathcal{X}%
_{b}^{[n]}$. If $g_{a}^{+}$ and $g_{b}^{+}$ are two tight upper guarantees
of respectively $\mathcal{W}_{a}$ and $\mathcal{W}_{b}$, then $%
g_{a}^{+}+g_{b}^{+}$ is clearly a tight guarantee of their sum\ $\mathcal{W}%
(x)=\mathcal{W}_{a}(x^{a})+\mathcal{W}_{b}(x^{b})$ on the domain $\mathcal{X}
$.

We do not know for which domain of functions $\mathcal{W}$ the converse
decentralisation property holds: \textit{every tight guarantee }$g$\textit{\
of} $\mathcal{W}_{a}+\mathcal{W}_{b}$\textit{\ is the sum of two tight
guarantees in the component problems.}

The answer eludes us even for the simple problem of assigning more than one
indivisible object and cash transfers when utilities over objects are
additive : the corresponding function $\mathcal{W}$ is the sum of problems $%
\mathcal{W}_{a}(x^{a})=\max_{i\in N}\{x_{i}^{a}\}$ over several objects $a$.
With much sweat we showed that the decentralisation property holds for two
agents and two objects!\footnote{%
The proof is available upon request from the authors.}\bigskip 

\textit{Acknowledgments: We are grateful for the comments we received in
WINE 2023 at Shanghai Tech University, SAET 2024 at the University of
Santiago, and Game Theory Days at UM6P Rabat; as well as in seminars at the
Universities of Paris 1, Vienna , Wuhan and Surrey. Special thanks to Dhivya
A. Kumar, Fedor Sandomirskiy, William Thomson, Erel Segal-Halevy, Miguel
Ballester, Yuqing Kong, John Quah and Bo Li.\medskip }

\pagebreak 

\section{Appendix: missing proofs}

\subsection{Statement $iii)$ in Lemma 3.1 and Lemma 3.2}

\textbf{\noindent }\textit{Step 1: A tight guarantee is
upper-hemi-continuous.} We fix $g^{-}\in \mathcal{G}^{--}$ and check that it
is u.h.c.. If it is not, there is in $\mathcal{X}$ some $x_{1}$, a sequence $%
\{x_{1}^{t}\}$ converging to $x_{1}$, and some $\delta >0$ such that $%
g^{-}(x_{1}^{t})\geq g^{-}(x_{1})+\delta $ for all $t$. Then we have, for
any $x_{-1}\in \mathcal{X}^{[n-1]}$: $\mathcal{W}(x_{1}^{t},x_{-1})\geq
g^{-}(x_{1}^{t})+\sum_{2}^{n}g^{-}(x_{i})\geq (g^{-}(x_{1})+\delta
)+\sum_{2}^{n}g^{-}(x_{i})$. Taking the limit in $t$ of $\mathcal{W}%
(x_{1}^{t},x_{-1})$ and ignoring the middle term we see that we can increase 
$g^{-}$ at $x_{1}$ without violating (\ref{3}), a contradiction of our
assumption $g^{-}\in \mathcal{G}^{--}$.\smallskip

\noindent \textit{Step 2: \textquotedblleft only if\textquotedblright\ in
Lemma 3.2}. We fix $g^{-}\in \mathcal{G}^{-}$ and show the existence of a
contact profile of $\mathcal{W}$ for any $x_{1}\in \mathcal{X}$. Define $%
\delta (x_{1})=\min_{x_{-1}\in \mathcal{X}^{N\diagdown 1}}\{\mathcal{W}%
(x_{1},x_{-1})-\sum_{N}g^{-}(x_{i})\}$ and note that this minimum is
achieved at some $\overline{x}_{-1}$ because the function $x_{-1}\rightarrow
\sum_{i=2}^{n}g^{-}(x_{i})$ is u.h.c. (step 1). Moreover $\delta (x_{1})$ is
non negative.

If $\delta (x_{1})=0$ then $(x_{1},\overline{x}_{-1})$ is a contact profile
of $\mathcal{W}$. If $\delta (x_{1})>0$ we can increase $g$ at $x_{1}$ to $%
g(x_{1})+\delta (x_{1})$, everything else equal, to get a guarantee
dominating $g^{-}$.\smallskip

\noindent \textit{Step 3: a tight guarantee is lower-hemi-continuous.} We
fix $g^{-}\in \mathcal{G}^{-}$ and check that it is l.h.c.. By the
continuity of $\mathcal{W}$ and compactness of $\mathcal{X}^{[n]}$ we have:%
\newline
$\forall \eta >0,\exists \theta >0,\forall x_{1},x_{1}^{\ast
},x_{-1}:||x_{1}-x_{1}^{\ast }||\leq \theta \Rightarrow \mathcal{W}%
(x_{1},x_{-1})\leq \mathcal{W}(x_{1}^{\ast },x_{-1})+\eta $

If $g^{-}$ is not l.h.c. there is some $x_{1}$ and $\{x_{1}^{t}\}$
converging to $x_{1}$ and $\delta >0$ s.t. $g^{-}(x_{1}^{t})\leq
g^{-}(x_{1})-\delta $ for all $t$. Pick $\theta $ for which the
approximation property above holds for $\eta =\frac{1}{2}\delta $; choose $t$
large enough that $||x_{1}^{t}-x_{1}||\leq \theta $: then for any $x_{-1}$: $%
g^{-}(x_{1})+\sum_{i=2}^{n}g^{-}(x_{i})\leq \mathcal{W}(x_{1},x_{-1})\leq 
\mathcal{W}(x_{1}^{t},x_{-1})+\frac{1}{2}\delta $. Replacing $g^{-}(x_{1})$
with $g^{-}(x_{1}^{t})+\delta \ $gives

$g^{-}(x_{1}^{t})+\sum_{i=2}^{n}g^{-}(x_{i})\leq \mathcal{W}%
(x_{1}^{t},x_{-1})-\frac{1}{2}\delta $ for any $x_{-1}$; this contradicts
the contact property of Lemma 3.2 for $x_{1}^{t}$.

\subsection{Statements $ii)$ and $iii)$ in Lemma 3.5}

\textbf{Proof }\textit{Statement }$iii)$ Fix $\varepsilon =-$, an arbitrary $%
\widetilde{x}_{1}\in \mathcal{X}$ and write $B(\widetilde{x}_{1},r)$\textbf{%
\ }for the closed ball of center $\widetilde{x}_{1}$\ and radius\textbf{\ }$%
r $. Use the notation $\Delta (x)=\sum_{1}^{n}una(x_{i})-\mathcal{W}(x)$ to
define the function $\delta (x_{1})=\max \{\Delta (x_{1},x_{-1}):\forall
i\geq 2,x_{i}\in B(\widetilde{x}_{1},d(x_{1},\widetilde{x}_{1}))\}$.

It is clearly continuous, non negative because $\Delta (x_{1},x_{-1})=0$ if $%
x_{i}=x_{1}$ for $i\geq 2$, and $\delta (\widetilde{x}_{1})=0$. Define $%
g=una-\delta $ and check that $g$ is a lower guarantee of $\mathcal{W}$. At
an arbitrary profile $x=(x_{i})_{1}^{n}$ choose $x_{i^{\ast }}$ s.t. $d(%
\widetilde{x}_{1},x_{i^{\ast }})$ is the largest: this implies $\delta
(x_{i^{\ast }})\geq \Delta (x)$. Combining this with $\delta (x_{i})\geq 0$
for $i\neq i^{\ast }$ gives $\sum_{1}^{n}\delta (x_{i})\geq \Delta (x)$
which, in turn, is the LH inequality in (\ref{3}) for $g$. As $g$ is in $%
\boldsymbol{G}^{-}$, it is dominated by some $\widetilde{g}$ in $\mathcal{G}%
^{-}$(Lemma 3.1) and $\widetilde{g}(x_{1})=una(\widetilde{x}_{1})$ (Lemma
3.5 $i)$).\smallskip

\noindent \textit{Second part of statement }$ii)$ We assume that $\mathcal{G}%
^{-}$ does not contain $una$ and check that $\mathcal{G}^{-}$ is not a
singleton. The assumption and the continuity of $\mathcal{W}$ imply that for
an open set of profiles $x\in \mathcal{X}^{[n]}$ we have $%
\sum_{[n]}una(x_{i})>\mathcal{W}(x)$. Fix such an $x$ and (by statement $%
iii) $) pick for each $i$ a tight guarantee $g_{i}^{-}$ equal to $una$ at $%
x_{i}$: these $n$ guarantees are not identical.

\subsection{Lemma 4.1}

Proof by contradiction\textit{.} Fix the even $n=2m$ and (without loss) $%
\mathcal{W}$ strictly supermodular. For some tight lower guarantee $g$ of $%
\mathcal{W}$ \textit{we have }$g(x_{i})=una(x_{i})$ at two types $x_{1}$\
and $x_{2}$ with $x_{1}<x_{2}$. This implies $mg(x_{1})+mg(x_{2})\leq 
\mathcal{W}(\overset{m}{x_{1}},\overset{m}{x_{2}})$ which we combine with $%
mg(x_{i})=\frac{1}{2}\mathcal{W}(\overset{n}{x_{i})}$ to get $\mathcal{W}(%
\overset{n}{x_{1}})+\mathcal{W}(\overset{n}{x_{2}})\leq 2\mathcal{W}(\overset%
{m}{x_{1}},\overset{m}{x_{2}})$. By repeated application of strict
supermodularity we have $\mathcal{W}(\overset{m}{x_{2}},\overset{m}{x_{2}})-%
\mathcal{W}(\overset{m}{x_{1}},\overset{m}{x_{2}})>\mathcal{W}(\overset{m}{%
x_{2}},\overset{m}{x_{1}})-\mathcal{W}(\overset{m}{x_{1}},\overset{m}{x_{1}}%
) $, precisely the opposite inequality. This is a contradiction.

The proof for odd $n=2m+1$ is similar. Summing up the inequality $%
(m+1)g(x_{1})+mg(x_{2})\leq \mathcal{W}(\overset{m+1}{x_{1}},\overset{m}{%
x_{2}})$ and the one exchanging $x_{1}$ and $x_{2}$ gives $\mathcal{W}(%
\overset{n}{x_{1}})+\mathcal{W}(\overset{n}{x_{2}})<\mathcal{W}(\overset{m+1}%
{x_{1}},\overset{m}{x_{2}})+\mathcal{W}(\overset{m}{x_{1}},\overset{m+1}{%
x_{2}})$ and we rewite the latter as another contradiction of strict
supermodularity.

\subsection{Theorem 4.1}

We fix $\mathcal{W}$ supermodular throughout and omit the routine
translation if it is submodular. In each statement the properties of $g_{L}$
are mirrored by properties of $g_{H}$ that we obtain by a similar argument
or, more elegantly, by applying statement $ii)$ of Lemma 3.4 tothe change of
variable $\theta (x_{i})=H+L-x_{i}$ exchanging $g_{L}$ and $g_{H}$.\smallskip

\textit{Statement }$i)$ Check that $g_{L}$ is a feasible lower guarantee. By
Remark 3.1 (just after Defnition 3.3) this implies that it is tight.
Consider the following inequality $\Pi _{q}$ for $q\in \lbrack n]$:%
\begin{equation*}
\mathcal{W}(x_{1},x_{2},\cdots ;x_{q},\overset{n-q}{L})+\sum_{\ell =q+1}^{n}%
\mathcal{W}(x_{\ell },\overset{n-1}{L})\leq \mathcal{W}(x)+(n-q)\mathcal{W}(%
\overset{n}{L})
\end{equation*}

Note that $\Pi _{n}$ is a tautology and $\Pi _{1}$: $\sum_{[n]}\mathcal{W}%
(x_{1};\overset{n-1}{L})\leq \mathcal{W}(x)+(n-1)\mathcal{W}(\overset{n}{L})$%
, means that $g_{L}$ is a lower guarantee (meets the LH of (\ref{3})). So it
is enough to prove inductively that $\Pi _{q+1}$ implies $\Pi _{q}$.

Supermodularity implies\newline
$\mathcal{W}(x_{q+1},\overset{n-1}{L})-\mathcal{W}(L,\overset{n-1}{L})\leq 
\mathcal{W}(x_{q+1},x_{1},\cdots ;x_{q},\overset{n-q-1}{L})-\mathcal{W}%
(L,x_{1},\cdots ;x_{q},\overset{n-q-1}{L})$. In the desired inequality $\Pi
_{q}$ we increase the left term $\mathcal{W}(x_{q+1},\overset{n-1}{L})$ by
the majoration we just obtained: this is now a weakly more demanding
inequality. After rearranging it we find the inequality $\Pi _{q+1}$ that we
assumed.\smallskip

\textit{Statement }$ii)$ If the tight guarantee $g^{-}$ in $\mathcal{G}^{-}$
satisfies $g^{-}(L\boldsymbol{)=}\frac{1}{n}(\overset{n}{L})$ we have for
all $x_{i}$: $g^{-}(x_{i})+(n-1)g^{-}(L)\leq \mathcal{W}(x_{i};\overset{n-1}{%
L})$ hence $g^{-}(x_{i})\leq g_{L}(x_{i})$. Therefore $g^{-}=g_{L}$ because $%
g^{-}$ is tight.\smallskip

\textit{Statement }$iii)$ Fix $g^{-}\in \mathcal{G}^{-}$ two types $%
x_{i},x_{i}^{\ast }$ s. t. $x_{i}<x_{i}^{\ast }$ and a contact profile $%
x_{-i}$ for $g^{-}$ at $x_{i}$; then apply the relevant inequality in Lemma
3.3:%
\begin{equation*}
g^{-}(x_{i}^{\ast })-g^{-}(x_{i})\leq \mathcal{W}(x_{i}^{\ast },x_{-i})-%
\mathcal{W}(x_{i},x_{-i})\leq \mathcal{W}(x_{i}^{\ast },\overset{n-1}{H})-%
\mathcal{W}(x_{i},\overset{n-1}{H})=g_{H}(x_{i}^{\ast })-g_{H}(x_{i})
\end{equation*}%
where the second inequality comes from the supermodularity of $\mathcal{W}$.

Next pick $x_{-i}^{\ast }$ a contact profile for $g^{-}$ at $x_{i}^{\ast }$
and apply again Lemma 3.3: $g^{-}(x_{i}^{\ast })-g^{-}(x_{i})\geq \mathcal{W}%
(x_{i}^{\ast },x_{-i}^{\ast })-\mathcal{W}(x_{i},x_{-i}^{\ast })\geq
g_{L}(x_{i}^{\ast })-g_{L}(x_{i})$.\smallskip

\textit{Statement }$iv)$ Suppose $g^{-}\in \mathcal{G}^{-}\diagdown
\{g_{L},g_{H}\}$ is s. t. $g^{-}(L)\leq g_{H}(L)$. We combine this
inequality with $g^{-}(x_{i})-g^{-}(L)\leq g_{H}(x_{i})-g_{H}(L)$ (statement 
$iii)$) and see that $g_{H}$ dominates $g^{-}$, a contradiction. So $%
g^{-}(L)>g_{H}(L)$. Next $g^{-}(L)\geq g_{L}(L)$ is ruled out by statement $%
ii)$ and $g^{-}\neq g_{L}$.,A similar argument proves the last two
inequalities. $\blacksquare $

\subsection{Proposition 4.2}

We prove the statement for the serial$\uparrow $\ rule (\ref{49}). By Lemma
3.6\textbf{\ }it is enough to check the inequality $g_{L}(x_{i})\leq \varphi
_{i}^{ser\uparrow }(x)\leq una(x_{i})$ for all $x$.

\textit{Step 1}. We show that $\varphi _{i}^{ser\uparrow }(x)$ increases
(weakly) in all variables $x_{j}$ such that $x_{j}\leq x_{i}$, i. e., for $%
j\leq i-1$ . This generalises Lemma 1 in \cite{MoSh1}.

If $\mathcal{W}$ is differentiable in $[L,H]^{n}$ we check this by computing
the derivative $\partial _{q}\varphi _{i}^{ser\uparrow }$ for $q\leq i-1$
in\ the LH of equation (\ref{49}) and using the symmetry of $\mathcal{W}$:%
\newline
$\partial _{q}\varphi _{i}^{ser\uparrow }(x)=\frac{\partial _{q}\mathcal{W}%
(x_{1},\cdots ,x_{i-1},\overset{n-i+1}{x_{i}})}{n-i+1}-\frac{\partial _{q}%
\mathcal{W}(x_{1},\cdots ,x_{q-1},\overset{n-q+1}{x_{q}})}{n-q}%
-\sum_{j=q+1}^{i-1}\frac{\partial _{q}\mathcal{W}(x_{1},\cdots ,x_{j-1},%
\overset{n-j+1}{x_{i}})}{(n-j+1)(n-j)}$. Recall that the coordinates of $x$
are weakly increasing. Because $\partial _{q}\mathcal{W}$ increases weakly
in $x_{j},j\neq q$, the numerator of each negative fraction is not larger
than that of the first fraction. The identity $\frac{1}{n-i+1}=\frac{1}{n-q}%
+\sum_{j=q+1}^{i-1}\frac{1}{(n-j+1)(n-j)}$ concludes the proof.\smallskip

Without the differentiability assumption the only step that requires an
additional argument is the following consequence of supermodularity: as the
coordinates of $x$ increase weakly the term $\mathcal{W}(x)-\frac{1}{n-q+1}%
\mathcal{W}(x_{1},\cdots ,x_{q-1},\overset{n-q+1}{x_{q}})$ increases weakly
in $x_{q}$ for each $q\leq n-1$. We omit the details.\smallskip

\textit{Step 2. }By construction of $\varphi ^{ser\uparrow }$ we have $%
\varphi _{i}^{ser\uparrow }(x)=\varphi _{i}^{ser\uparrow }(x_{1},\cdots
,x_{i-1},\overset{n-i+1}{x_{i}})$ and by Step 1 it is enough to check that $%
g_{L}(x_{i})$ lower bounds $\varphi _{i}^{ser\uparrow }(x)$ at the profile $(%
\overset{i-1}{L},\overset{n-i+1}{x_{i}})$ while $una$ upper bounds it at $(%
\overset{n}{x_{i}})$. The latter follows from $\varphi _{i}^{ser\uparrow }(%
\overset{n}{x_{i}})=una(x_{i})$.

Applying (\ref{49}) we see that $g_{L}\leq \varphi _{i}^{ser\uparrow }$
reduces to $\mathcal{W}(\overset{n-1}{L},x_{i})\leq \frac{1}{n-i+1}\mathcal{W%
}(\overset{i-1}{L},\overset{n-i+1}{x_{i}})+\frac{n-i}{n-i+1}\mathcal{W}(%
\overset{n}{L})$ or equivalently $(n-i)(\mathcal{W}(\overset{n-1}{L},x_{i})-%
\mathcal{W}(\overset{n}{L}))\leq \mathcal{W}(\overset{i-1}{L},\overset{n-i+1}%
{x_{i}})-\mathcal{W}(\overset{n-1}{L},x_{i})$.

Finally we apply supermodularity to successively lower bound $\mathcal{W}(%
\overset{q}{L},\overset{n-q}{x_{i}})-\mathcal{W}(\overset{q+1}{L},\overset{%
n-q-1}{x_{i}})$ by $\mathcal{W}(\overset{n-1}{L},x_{i})-\mathcal{W}(\overset{%
n}{L})$ for $q=(n-2),\cdots ,(i-1)$ and sum up these inequalities.

\subsection{Proposition 5.1}

We assume without loss that $F$ is convex.

\noindent \textit{Statement }$i)$ \textit{We show that }$g_{\ell ,h}$\textit{%
\ defined in the Proposition is a lower guarantee: }$g_{\ell ,h}\in \mathbf{G%
}^{-}$. By Remark 3.1 this implies that $g_{\ell ,h}$ is tight.

We set $Z=\ell L+hH$ for easier reading. The feasibility inequality (\ref{3}%
) applied to $g_{\ell ,h}$ reads%
\begin{equation}
\sum_{\lbrack n]}F(x_{i}+Z)\leq F({\large x}_{N})+\ell F(Z+L)+hF(Z+H)\text{
for }x\in \lbrack L,H]^{[n]}  \label{47}
\end{equation}

We proceed by induction on $n$. There is nothing to prove if $n=2$. For $n=3$
we already know that $g_{2,0}$ and $g_{0,2}$ are in $\mathcal{G}^{-}$; for $%
g_{1,1}$ the inequality above is $\sum_{i=1}^{3}F(x_{i}+L+H)\leq
F(x_{123})+F(2L+H)+F(L+2H)$.

Suppose $x_{12}\geq L+H$; the convexity of $F$ gives $F(x_{3}+L+H)-F(2L+H)%
\leq F(x_{123})-F(x_{12}+L)$. Replacing $F(x_{3}+L+H)$ in the desired
inequality by this upper bound and rearranging gives a more demanding
inequality: $F(x_{1}+L+H)+F(x_{2}+L+H)\leq F(x_{12}+L)+F(L+2H)$. It follows
again from the convexity of $F$ (because $x_{1}+H,x_{2}+H\in \lbrack
x_{12},2H]$. So we are done if $x_{ij}\geq L+H$ for some pair $i,j$.

Suppose next $x_{ij}\leq L+H$ for all three pairs. Then $x_{123},2L+H\leq
x_{i}+L+H\leq L+2H$ for $i=1,2,3$. And the uniform distribution on the
triple $x_{123},2L+H,L+2H$ is a mean-preserving spread of that on $%
(x_{i}+L+H)_{i=1}^{3}$, as was to be proven.\smallskip

For the inductive argument we fix $n\geq 4$ and $g_{\ell ,h}$ s. t. $\ell
+h=n-1$ and $\ell \geq 1$. We assume that (\ref{47}) holds for $n-1$ agent
problems and prove it for $(\ell ,h)$.

Suppose $x_{N\diagdown \{n\}}\geq Z$ for some agent labeled $n$ without loss
of generality. Then the convexity of $F$ implies $F(x_{n}+Z)-F(Z+L)\leq
F(x_{N})-F(x_{N\diagdown \{n\}}+L)$.

As before we replace $F(x_{n}+Z)$ by this upper bound and rearrange (\ref{47}%
) to the more demanding $\sum_{[n-1]}F(x_{i}+Z)\leq F(x_{N\diagdown
\{n\}}+L)+(\ell -1)F(Z+L)+hF(Z+H)$, which for the convex function $%
\widetilde{F}(y)=F(y+L)$ and $\widetilde{Z}=(\ell -1)L+hH$ is exactly (\ref%
{47}) at $x_{-n}$ for the guarantee $g_{(\ell -1),h}$.

We are left with the case where $x_{N\diagdown \{i\}}\leq Z$ for all $i$ for
which the different terms under $F$ in (\ref{47}) are ranked as follows: $%
x_{N}$ and $Z+L\leq x_{i}+Z\leq Z+H$. The distribution $(\frac{1}{n},\frac{%
\ell }{n},\frac{h}{n})$ on the support $x,Z+L,Z+H$ is a mean-preserving
spread of the uniform distribution on the $n$ inputs $x_{i}+Z$. So $g_{\ell
,h}$ meets (\ref{47}).

We still have to deal with the case $\ell =0,h=n-1$. Then $g_{0,h}$ is just
the canonical SA $g_{H}$ and we are done.\smallskip

\noindent \textit{Statement }$ii)$ The derivative of the gap function is $%
\frac{dF}{dx}(nx_{i})-\frac{dF}{dx}(x_{i}+Z)$ which changes from negative to
positive at $x_{i}=\frac{1}{n-1}Z$ where it achieves the smallest gap. The
equality $g_{\ell ,h}(Z)=una(Z)$ is rearranged as: $F(Z)=\frac{1}{n}F(nZ)+%
\frac{\ell }{n}F(Z+L)+\frac{h}{n}F(Z+H){\large )}$. This contradicts the
strict convexity of $F$ if $\ell ,h$ are both positive.

\subsection{Proposition 5.2}

We assume without loss that $F$ is convex so that $\mathcal{W}$ is
supermodular.

\noindent \textit{Case 1} We already noted that $\theta _{a}$ is in $\mathbf{%
G}^{-}$. For tightness we fix a type $x_{i}$ and look for a vector $x_{-i}$
such that $x_{i}+x_{N\diagdown i}=na$. This implies $\sum_{[n]}\theta
_{a}(x_{j})=F(na)$ by the definition of $\theta _{a}$, so $(x_{i},x_{-i})$
is a contact profile of $\theta _{a}$ at $x_{i}$ and we are done by Lemma
3.2. The desired vector $x_{-i}$ exists if and only if $x_{i}+(n-1)L\leq
na\leq x_{i}+(n-1)H$, precisely as we assume.\smallskip

\noindent \textit{Case 2} At a profile $x$ where $x_{i}\leq na-(n-1)L$ for
all $i$, we just saw that $g_{a}=\theta _{a}$ meets the LH of (\ref{3}). We
check now this inequality for a profile $x$ where the first $t$ types are
above $na-(n-1)L$, $t\geq 1$, and the other $n-t$ types (possibly zero) are
below that bound. For $i\leq t$\ we can write $%
g_{a}(x_{i})=F(x_{i}+(n-1)L)+C_{i}$ where $C_{i}$ is a constant w r t $x$,
and similarly $g_{a}(x_{j})=\partial \theta _{a}\times x_{j}+C_{j}$ if $j>t$%
. Then the desired LH inequality of (\ref{3}) is%
\begin{equation}
\sum_{i\leq t}F(x_{i}+(n-1)L)+\sum_{j>t}\partial \theta _{a}\times
x_{j}+C\leq F(x_{N})  \label{34}
\end{equation}%
for some constant $C$.

For $i\leq t$ the difference $F(x_{i}+x_{N\diagdown i})-F(x_{i}+(n-1)L)$ is
smallest for $x_{i}=na-(n-1)L$ so it is enough to prove (\ref{34}) if this
is the case. For $j>t$ we check similarly that the difference $\Delta
=F(x_{j}+x_{N\diagdown j})-\partial \theta _{a}\times x_{j}$ is smallest if $%
x_{j}=L$. Note that $t\geq 1$ implies $x_{N\diagdown j}\geq
na-(n-1)L+(n-2)L=na-L$, therefore the derivative of $F(x_{j}+x_{N\diagdown
j})$ w r t $x_{j}$ at any $x_{j}>L$ is weakly larger than $\partial \theta
_{a}$ which proves the claim. So it is enough to prove (\ref{34}) if $%
x_{j}=L $ for $j>t$ and $x_{i}=na-(n-1)L$ for $i\leq t$. In this case we
have $x_{N}=tna-(t-1)nL$, $x_{N}\geq na$ and (\ref{34}) is $%
tg_{a}(na-(n-1)L)+(n-t)g_{a}(L)\leq F(tna-(t-1)nL)$.

Equivalently $\partial \theta _{a}\times (t-1)n(a-L)\leq
F(tna-(t-1)nL)-F(na) $, a consequence of the convexity of $F$.

Check tightness. At type $x_{i}\leq na-(n-1)L$ we have $x_{i}+(n-1)L\leq
na\leq x_{i}+(n-1)(na-(n-1)L)$ \ by replacing $x_{i}$ by $L$ in the last
term and rearranging. This implies the existence of a profile $x_{-i}$
entirely inside $[L,na-(n-1)L]$ and s. t. $x_{i}+x_{N\diagdown i}=na$: as in
Case 1 this is a contact profile. And at a type $x_{i}\geq na-(n-1)L$ the
definition of $g_{a}$ shows directly that $(x_{i},\overset{n-1}{L})$ is a
contact profile.

We omit the symmetric proof of Case 3. $\blacksquare $

\subsection{Lemma 6.1}

Fix the continuous functions $w_{q}$ and $\mathcal{W}$ as in Definition 6.1.

\textit{Proof of \textquotedblleft only if\textquotedblright\ }We assume
that $\mathcal{W}$ is supermodular and apply repeatedly Definition 4.1 in
the form of inequality (\ref{38}). For any $(n-2)$-profile $x_{-12}$ and
4-tuple of types s. t. $y_{1}<x_{1}$ and $y_{2}<x_{2}$ we have $\mathcal{W}%
(x_{1},x_{2};x_{-12})-\mathcal{W}(y_{1},x_{2};x_{-12})\geq \mathcal{W}%
(x_{1},y_{2};x_{-12})-\mathcal{W}(y_{1},y_{2};x_{-12})$.

Fix $x_{1},y_{1}$ s. t. $L<y_{1}<x_{1}<H$ and pick any rank $q$ except $n$.
If we pick $y_{2},x_{2}$ so that $L<y_{2}<y_{1}<x_{1}<x_{2}<H$ then we can
choose $x_{-12}$ so that $x_{1}$ and $y_{1}$ are of rank $q$ in the profiles
on the RH of (\ref{38}), whereas after increasing $y_{2}$ to $x_{2}$ they
are of rank $q+1$ in the profiles on the LH. Then (\ref{38}) amounts to $%
w_{q+1}(x_{1})-w_{q+1}(y_{1})\geq w_{q}(x_{1})-w_{q}(y_{1})$. Because $%
\mathcal{W}$ is continuous this desired inequality holds for any $%
x_{1},y_{1} $ s. t. $x_{1}\leq y_{1}$.\smallskip

\textit{Proof of \textquotedblleft if\textquotedblright }\ We are given the
continuous $w_{q}$ such that $w_{q}$\textit{\ }grows weakly slower than $%
w_{q+1}$ for all $q\leq n-1$. We show that the rank-separable function $%
\mathcal{W}$ is supermodular, i. e. we prove (\ref{38}) for any $%
x,y_{1},y_{2}$ s. t. $y_{1}\leq x_{1}$ and $y_{2}\leq x_{2}$. As $\mathcal{W}
$ is continuous it is enough to prove it when $y_{1},y_{2}$ and all the
coordinates of $x$ are diffferent.

Next it is without loss to assume that in the jump up from $y_{i}$ to $x_{i}$
either the rank of this coordinate does not change, or it goes down but
exactly one. Indeed if $y_{1}$ is ranked $q$ in $(y_{1},x_{2};x_{-12})$ and $%
x_{1}$ ranked $q-3$ in $(x_{1},x_{2};x_{-12})$ we decompose the jump in
three small jumps each decreasing the rank by one, and sum up the
corresponding inequalities (\ref{38}). The same argument applies to the jump
from $y_{2}$ to $x_{2}$.

In the profile $(y_{1},y_{2};x_{-12})$ we call $q_{i}$ the rank of $y_{i}$,
so $q_{1}>q_{2}$. For clarity we omit the fixed term $x_{-12}$ in the
computations below and define $\Delta _{1}=\mathcal{W}(x_{1},y_{2})-\mathcal{%
W}(y_{1},y_{2})$ ; $\Delta _{2}=\mathcal{W}(y_{1},x_{2})-\mathcal{W}%
(y_{1},y_{2})$ ; $\Delta _{0}=\mathcal{W}(x_{1},x_{2})-\mathcal{W}%
(y_{1},y_{2})$, so that (\ref{38}) is equivalent to $\Delta _{0}\geq \Delta
_{1}+\Delta _{2}$.

We assume $y_{1}<y_{2}$ without loss and consider first the case $%
x_{1}<y_{2} $. Then $\Delta _{1}=w_{q_{1}}(x_{1})-w_{q_{1}}(y_{1})$ if agent 
$1$'s rank does not change in the move from $y_{1}$ to $x_{1}$; or $\Delta
_{1}=w_{q_{1}-1}(x_{1})+w_{q_{1}}(x_{j})-w_{q_{1}-1}(x_{j})-w_{q_{1}}(y_{1})$
if agent $1$ jumps above agent $j$ of rank $q_{1}-1$ (that cannot be agent $%
2 $). Next $\Delta _{2}=w_{q_{1}}(x_{2})-w_{q_{1}}(y_{2})$ and we see that $%
\Delta _{0}=\Delta _{1}+\Delta _{2}$.

Assume from now on $x_{1}>y_{2}$ which implies $q_{1}=q_{2}+1$, and agent $1$%
's rank at $(x_{1},y_{2})$ is $q_{2}$ (because each rank upgrade is at most
one). We distinguish three cases.

If $y_{1}<y_{2}<x_{2}<x_{1}$ then only agents $1$ and $2$ swap ranks and $%
\Delta _{2}=w_{q_{2}}(x_{2})-w_{q_{2}}(y_{2})$; $\Delta
_{1}=w_{q_{2}}(x_{1})+w_{q_{2}+1}(y_{2})-w_{q_{2}}(y_{2})-w_{q_{2}+1}(y_{1})$%
; $\Delta
_{0}=w_{q_{2}}(x_{1})+w_{q_{2}+1}(x_{2})-w_{q_{2}}(y_{2})-w_{q_{2}+1}(y_{1})$
so that $\Delta _{0}-\Delta _{1}+\Delta
_{2}=(w_{q_{2}+1}(x_{2})-w_{q_{2}+1}(y_{2}))-(w_{q_{2}}(x_{2})-w_{q_{2}}(y_{2})) 
$, which is non negative by assumption.

If $y_{1}<y_{2}<x_{1}<x_{2}$ and agent $2$'s rank at $x_{2}$ is still $q_{2}$%
, then $\Delta _{1}$ and $\Delta _{2}$ are just as in the previous case, and 
$\Delta
_{0}=w_{q_{2}}(x_{2})+w_{q_{2}+1}(x_{1})-w_{q_{2}}(y_{2})-w_{q_{2+1}}(y_{1})$
implies $\Delta _{0}-\Delta _{1}+\Delta
_{2}=(w_{q_{2}+1}(x_{1})-w_{q_{2}+1}(y_{2}))-(w_{q_{2}}(x_{1})-w_{q_{2}}(y_{2})) 
$ and the conclusion.

We omit for brevity the similar computations of the last case where agent $2$%
's rank at $x_{2}$ is $q_{2}-1$.

\subsection{Theorem 6.1}

\subsubsection{Statement $i)$ $\Rightarrow $ statement $ii)$}

We fix $\mathcal{W}$ rank-separable and supermodular.

\textbf{Step 1}.\textit{For any }$c$\textit{\ the function }$g_{c}$\textit{\
defined by (\ref{22}) is in }$\mathcal{G}^{-}$. By Remark 3.1 it is enough
to show $g_{c}\in \mathbf{G}^{-}$.

Because $g_{c}(x_{i})$ and $\mathcal{W}(x_{i};c)$ are continuous in $x_{i},c$
it is enough to prove the LH inequality (\ref{3}) for strictly decreasing
sequences $\{x_{\ell }\}_{1}^{n}$ and $\{c_{q}\}_{1}^{n-1}$ s. t. $H>c_{1}$
and $c_{n-1}>L$ and $x_{\ell }\neq c_{q}$ for all $\ell ,q$. This is always
assumed in the rest of the proof.\smallskip

\textit{Step 1.1} Call the ordered sequence of types $x$ \textit{regular (w
r t }$c$) if $x_{1}>c_{1}>x_{2}>c_{2}>\cdots >c_{q-1}>x_{q}>c_{q}>\cdots
>c_{n-1}>x_{n}$. Then check that $x$ is a contact profile of $g_{c}$: $%
\sum_{1}^{n}g_{c}(x_{q})=\sum_{1}^{n}\mathcal{W}(x_{q},c)-\sum_{1}^{n-1}%
\mathcal{W}(c_{q},c)=\sum_{1}^{n-1}(w_{q}(x_{q})-w_{q}(c_{q}))+\mathcal{W}%
(x_{n},c)=\mathcal{W}(x)$.

Recall that $(x_{i},c)$ is also a contact profile (Definition 3.3).\smallskip

\textit{Step 1.2} For any three sequences $x,x^{\prime }$ and $c$ we say
that $x^{\prime }$ is reached from $x$ by an \textit{elementary jump up
above }$c_{q}$ if there is some $\ell $ such that $x_{-\ell }=x_{-\ell
}^{\prime }$; $c_{q}$ is adjacent to $x_{\ell }$ in $x$ from above and
adjacent to $x_{\ell }^{\prime }$ in $x^{\prime }$ from below. In other
words $x_{\ell }^{\prime }>c_{q}>x_{\ell }$ and there is no other element of 
$x$ or $c$ between $x_{\ell }$ and $x_{\ell }^{\prime }$. The definition of
an elementary jump down below $c_{q}$ is symmetrical.

Starting from for an arbitrary profile $\widetilde{x}$ we construct the
canonical path of profiles $\sigma =\{\widetilde{x}=$ $^{1}x,\cdots ,^{\ell
}x,\cdots ,^{T}x=x^{\ast }\}$ from $\widetilde{x}$ to a regular profile $%
x^{\ast }$ such that 1) each step from $^{\ell }x$ to $^{\ell +1}x$ is an
elementary jump up or down of some $^{t}x_{\ell }$ over some $c_{q}$ and 2) $%
\ell \leq q$ if $^{t}x_{\ell }$ jumps up above $c_{q}$, and $\ell \geq q+1$
if $^{t}x_{\ell }$ jumps down below $c_{q}$.

Case 1: $\widetilde{x}_{1}>c_{1}$. Then $\widetilde{x}_{1}$ never moves and $%
\widetilde{x}_{1}=x_{1}^{\ast }$; if $\widetilde{x}_{2},\cdots ,\widetilde{x}%
_{\ell }$ are above $c_{1}$ then $\ell -1$ successive elementary jumps down
of $\widetilde{x}_{\ell }$, then $\widetilde{x}_{\ell +1}$, etc.. below $%
c_{1}$ defines the first $\ell -1$ steps of the desired path, and we are
left with the shorter sequences $\widetilde{x}_{-1}$ and $c_{-1}$.

Case 2: $c_{1}>\widetilde{x}_{1}$. Then the successive elementary jumps up
of $\widetilde{x}_{1}$ over the closest $c_{q}$ then $c_{q-1},\cdots ,c_{1}$
define the first $q$ steps of the desired path until $^{q+1}x=x_{1}^{\ast }$
that never moves again; then as above we use the induction for the sequences 
$\widetilde{x}_{-1}$ and $c_{-1}$.\smallskip

\textit{Step 1.3} We pick an arbitrary profile $\widetilde{x}$, construct a
sequence $\sigma $ from $\widetilde{x}$ to some regular $x^{\ast }$, and
check that in each step of the sequence the sum $\sum_{1}^{n}g_{c}(x_{\ell
})-\mathcal{W}(x)$ cannot decrease, which together with Step 1.1 concludes
the proof that $g_{c}\in \mathbf{G}^{-}$. This sum develops as $\overset{B}{%
\overbrace{(\sum_{\ell =1}^{n}\mathcal{W}(x_{\ell },c))}}-\overset{C}{%
\overbrace{\mathcal{W}(x)}}-\overset{D}{\overbrace{\sum_{q=1}^{n-1}\mathcal{W%
}(c_{q},c)}}$. Fix an elementary jump up of $^{t}x_{\ell }$ above $c_{q}$: $%
^{t+1}x_{\ell }>c_{q}>$ $^{t}x_{\ell }$. The net changes to the three terms
in the sum are $\Delta B=w_{q}(^{t+1}x_{\ell })-w_{q+1}(^{t}x_{\ell
})+w_{q+1}(c_{q})-w_{q}(c_{q})$; $\Delta C=w_{\ell }(^{t+1}x_{\ell
})-w_{\ell }(^{t}x_{\ell })$ ; $\Delta D=0$.

With the notation $\Delta (f;a\rightarrow b)=f(b)-f(a)$ and some rearranging
this gives $\Delta B-\Delta C+\Delta D=\Delta (w_{q}-w_{\ell
};c_{q}\rightarrow $ $^{t+1}x_{\ell })+\Delta (w_{q+1}-w_{\ell };^{t}x_{\ell
}\rightarrow c_{q})$, where both final $\Delta $ terms are non negative
because $\ell \leq q$ and by Lemma 6.1 $w_{q}-w_{\ell }$ and $%
w_{q+1}-w_{\ell }$ increase weakly.

The proof for an elementary jump down is quite similar by computing the
variation of\newline
$\sum_{1}^{n}g_{c}(x_{\ell })-\mathcal{W}(x)$ to be $\Delta (w_{\ell
}-w_{q})(c_{q}\rightarrow x_{\ell }^{t})+\Delta (w_{\ell }-w_{q+1})(x_{\ell
}^{t+1}\rightarrow c_{q})$ and recalling that in this case we have $\ell
\geq q+1$.\smallskip

\textbf{Step 2} \textit{A tight guarantee} $g\in \mathcal{G}^{-}$ \textit{of}
$\mathcal{W}$\textit{\ takes the form }$g_{c}$\textit{\ in (\ref{22})
(Definition 3.3).}

Recall the notation $\mathcal{C}(g)$ for the set of contact profiles of $g$
(Lemma 3.2). For each $q\in \lbrack n]$ its projection $\mathcal{C}_{q}(g)$
is the set of those $x_{i}\in \lbrack L,H]$ appearing in some profile $x\in 
\mathcal{C}(g)$ with the rank $q$; it is closed because $\mathcal{C}(g)$ is
closed and we call its lower bound $c_{q}$. The sequence $\{c_{q}\}$
decreases weakly because in a contact profile where $c_{q}$ has rank $q$ the
type $x_{q+1}$ ranked $q+1$ is below $c_{q}$. Also $c_{n}=L$ because $g$ is
tight so $c_{n}$ is in one of its contact profiles.

We show first that $\mathcal{C}_{1}(g)=[c_{1},H]$ and that in this interval $%
g$ \textquotedblleft follows\textquotedblright\ $w_{1}$, i. e., $g-w_{1}$ is
a constant. The critical tool is Lemma 3.3 and we also keep in mind that $g$
is continuous (Lemma 3.1). Pick a profile $\overline{x}\in \mathcal{C}(g)$
s. t. $\overline{x}^{1}=c_{1}$ and apply Lemma 3.3 to $c_{1}$ and an
arbitrary $\widehat{x}_{1}$ in $[c_{1},H]$ to get $g(\widehat{x}%
_{1})-g(c_{1})\geq w_{1}(\widehat{x}_{1})-w_{1}(c_{1})$. Combine this with
the contact equation for $\overline{x}$: $g(c_{1})-w_{1}(c_{1})=%
\sum_{2}^{n}(w_{q}(\overline{x}_{q})-g(\overline{x}_{q}))\leq g(\widehat{x}%
_{1})-w_{1}(\widehat{x}_{1})$ and recall that $g$ is a lower guarantee: the
latter inequality must be an equality therefore $\widehat{x}_{1}$ is in $%
[c_{1},H]$; as $\widehat{x}_{1}$ was arbitrary this shows $[c_{1},H]=%
\mathcal{C}_{1}(g)$ and that $g-w_{1}$ is a constant in $[c_{1},H]$.

We show next that $[c_{2},c_{1}]\subseteq \mathcal{C}_{2}(g)$ and $g$
follows\ $w_{2}$ in this interval. Pick any $\widehat{x}_{2}\in \lbrack
c_{2},c_{1}[$ and $\overline{x}$ $\in \mathcal{C}(g)$ s. t. $\overline{x}%
^{2}=c_{2}$ and apply again Lemma 3.3 to $c_{2}$ and $\overline{x}$:%
\begin{equation}
g(\widehat{x}_{2})-g(c_{2})\geq \mathcal{W}(\widehat{x}_{2},\overline{x}%
_{-2})-\mathcal{W}(c_{2},\overline{x}_{-2})  \label{42}
\end{equation}

The rank of $\widehat{x}_{2}$ in $(\widehat{x}_{2},\overline{x}_{-2})$ is at
least 2 because $\widehat{x}_{2}<c_{1}$. If it is 2 the RH term above is $%
w_{2}(\widehat{x}_{2})-w_{2}(c_{2})$. We combine again this inequality with
the contact equation for $c_{2}$ to get $\sum_{q\neq 2}(w_{q}(\overline{x}%
_{q})-g(\overline{x}_{q}))\leq g(\widehat{x}_{2})-w_{2}(\widehat{x}_{2})$
and deduce exactly as before that $\overline{x}_{-2}$ is a contact profile
of $\widehat{x}_{2}$ so that $\widehat{x}_{2}\in \mathcal{C}_{2}(g)$
moreover $g-w_{2}$ is constant for those types $\widehat{x}_{2}$ in $%
[c_{2},c_{1}]$ of rank 2 in $(\widehat{x}_{2},\overline{x}_{-2})$.

If the rank of $\widehat{x}_{2}$ in $(\widehat{x}_{2},\overline{x}_{-2})$ is
3, inequality (\ref{42}) becomes $g(\widehat{x}_{2})-g(c_{2})\geq w_{2}(%
\overline{x}_{3})+w_{3}(\widehat{x}_{2})-w_{2}(c_{2})-w_{3}(\overline{x}%
_{3}) $. After using the contact equation of $(c_{2},\overline{x}_{-2})$ to
replace $w_{2}(c_{2})-g(c_{2})$ by $\sum_{q\neq 2}(g(\overline{x}_{q})-w_{q}(%
\overline{x}_{q}))$ we obtain $g(\widehat{x}_{2})+\sum_{q\neq 2}g(\overline{x%
}_{q})\geq (\sum_{q\neq 2,3}w_{q}(\overline{x}_{q}))+w_{2}(\overline{x}%
_{3})+w_{3}(\widehat{x}_{2})$. Because $g$ is a lower guarantee this is an
equality, proving that $(\widehat{x}_{2},\overline{x}_{-2})\in \mathcal{C}%
(g) $ and $\widehat{x}_{2}\in \mathcal{C}_{2}(g)$; moreover $g-w_{2}$.is
constant for this subset of types in $[c_{2},c_{1}]$.

The similar argument when the rank $q$ of $\widehat{x}_{2}$ in $(\widehat{x}%
_{2},\overline{x}_{-2})$ is larger than 3 should now be clear: it concludes
the proof that $[c_{2},c_{1}]\subseteq \mathcal{C}_{2}(g)$, moreover
identifies at most $n-q+1$ closed subsets of the interval in which $g-w_{2}$%
.is constant: this function is continuous therefore it is constant on the
whole interval.

There is no additional difficulty to prove for all $q\in \lbrack n]$ the
inclusion $[c_{q},c_{q-1}]\subseteq \mathcal{C}_{q}(g)$ and the fact that $%
g-w_{q}$ is constant in this interval. Now the guarantee $g_{c}$ for $%
c=(c_{q})_{q=1}^{n-1}$ also follows $w_{q}$ as well in $[c_{q},c_{q-1}]$:
both $g$ and $g_{c}$ are continous so they differ by a constant, that must
be zero because they are both tight.

\subsubsection{Statement\textbf{\ }$ii)$ $\Rightarrow $ statement $i)$}

Fixing $\mathcal{W}$ supermodular and s. t. each function $g_{c},c\in
\lbrack L.H]^{n-1}$ is a tight lower guarantee of $\mathcal{W}$, we prove\
that $\mathcal{W}$ is rank-separable. If in the pair $x,c$ the profile $x$
is regular w r t $c$ \textit{\ (Step 1.1) we simply say that the pair }$x,c$
is regular.

\textbf{Step 1 }\textit{For any regular pair }$x,c$\textit{\ we have}%
\begin{equation}
\mathcal{W}(x)=\sum_{q=1}^{n}\mathcal{W}(x_{q},c)-\sum_{\ell =1}^{n-1}%
\mathcal{W}(c_{\ell },c)  \label{41}
\end{equation}

By the Definition 3.3 of $g_{c}$ ((\ref{22})) the RH term of (\ref{41}) is
just $\sum_{q=1}^{n}g_{c}(x_{q})$ hence bounded above by $\mathcal{W}(x)$ by
assumption. We check the opposite inequality by induction on $n$.

For $n=2$ the supermodularity and symmetry of $\mathcal{W}$ imply at once $%
\mathcal{W}(x_{1},x_{2})+\mathcal{W}(c,c)\leq \mathcal{W}(x_{^{1}},c)+%
\mathcal{W}(x_{^{2}},c)$ whenever $x_{1}\geq c\geq x_{2}$. For a general $n$
repeated application of supermodularity implies $\mathcal{W}(c_{n-1},c)-%
\mathcal{W}(x_{n},c)\leq \mathcal{W}(c_{n-1},x_{-n})-\mathcal{W}%
(x_{n},x_{-n})$. We strengthen the desired inequality in (\ref{41}) by
adding to its LH the RH of this last inequality, and its LH to the RH of (%
\ref{41}). This operation produces, after rearranging, the inequality\newline
$\mathcal{W}(c_{n-1},x_{-n})\leq \sum_{q=1}^{n-1}\mathcal{W}%
(x_{q},c)-\sum_{\ell =1}^{n-2}\mathcal{W}(c_{\ell },c)$, precisely (\ref{41}%
) for the supermodular symmetric function of $(n-1)$ variables $%
x_{-n}\rightarrow \mathcal{W}^{\blacklozenge }(x_{-n})=\mathcal{W}%
(c_{n-1},x_{-n})$.\smallskip

\textbf{Step 2 }\textit{Solving the functional equation}

For $n=2$ we apply (\ref{41}) to $x_{1}\geq c\geq x_{2}=L$: $\mathcal{W}%
(x_{1},L)=\mathcal{W}(x_{1},c)+\mathcal{W}(L,c)-\mathcal{W}%
(c,c)\Longleftrightarrow \mathcal{W}(x_{1},c)=\mathcal{W}(x_{1},L)+w_{2}(c)$
where $w_{2}(c)=\mathcal{W}(c,c)-\mathcal{W}(L,c)$. As $x_{1}$ and $c$ can
be freely chosen in $[L.H]$ provided $x_{1}\geq c$, the rank separability of 
$\mathcal{W}$ follows.

Assuming the result is true until $n-1$ we apply (\ref{41}) for $\mathcal{W}$
with domain $[L,H]^{n}$ to a regular pair $x,c$ s. t. $x_{n}=c_{n-1}=L$: $%
\mathcal{W}(x_{-n},L)=\{\sum_{q=1}^{n-1}\mathcal{W}(x_{q},c_{-(n-1)},L)\}+%
\mathcal{W}(c_{-(n-1)},\overset{2}{L})-\{\sum_{\ell =1}^{n-2}\mathcal{W}%
(c_{\ell },c_{-(n-1)},L)\}-\mathcal{W}(c_{-(n-1)},\overset{2}{L})$.

For the function $\mathcal{W}^{\blacklozenge }(x_{-n})=\mathcal{W}(x_{-n},L)$
over $[L,H]^{n-1}$ this is exactly equation (\ref{41}) so by the inductive
assumption $\mathcal{W}^{\blacklozenge }$ is rank-separable. For some
continuous functions $\theta _{1},\cdots ,\theta _{n-1}$ on $[L,H]$ we have $%
\mathcal{W}(x_{-n},L)=\sum_{q=1}^{n-1}\theta _{q}(x_{q})$.for any decreasing
sequence $x_{1},\cdots ,x_{n-1}$.

We now apply (\ref{41}) to the regular pair $x,c$ such that $x_{q}=c_{q}$
for $2\leq q\leq n-1$, and $x_{n}=L$: $\mathcal{W}(x_{1},c_{-1},L)=\mathcal{W%
}(x_{1},c)+\sum_{q=2}^{n-1}\mathcal{W}(c_{q},c)+\mathcal{W}(c,L)-\sum_{\ell
=1}^{n-1}\mathcal{W}(c_{\ell },c)$. Taking advantage of separability of $%
\mathcal{W}(\cdot ,L)$ in the first $n-1$ variables, this reduces to: $%
\theta _{1}(x_{1})-\theta _{1}(c_{1})=\mathcal{W}(x_{1},c)-\mathcal{W}%
(c_{1},c)$.

In this choice of $x,c$ the weakly decreasing sequence $x_{1},c_{1},\cdots
,c_{n-1}$ is arbitrary. Therefore $\mathcal{W}$ separates its largest
variable from the $n-1$ others: for some continous functions $\tau ,T$ we
have $\mathcal{W}(x)=\tau (x_{1})+T(x_{-1})$ if $x_{1}$ is a largest
coordinate.

Just as we showed that $\mathcal{W}(x_{-n},L)$ is rank-separable by the
inductive assumption, we see now (by applying (\ref{41}) when $x_{1}=c_{1}=H$%
) that $\mathcal{W}(H,x_{-1})$ is rank-separable too: $\mathcal{W}%
(H,x_{-1})=\sum_{q=2}^{n}\lambda _{q}(x_{q})$ for some continuous $\lambda $%
-s. We also know $\mathcal{W}(H,x_{-1})=\tau (H)+T(x_{-1})$ therefore $T$ is
additively separable as well over decreasing sequences and we are done.

\subsection{Lemma 7.2}

\textit{Statement} $i)$ is clear because $\mathcal{W}$ is symmetric. In 
\textit{Statement} $ii)$ upper-hemi-continuity of $\gamma $ is clear because 
$\mathcal{W}$ and $g$ are both continous (step 1 in the proof of Lemma 3.1,
Section 9.1).

To check that $\gamma $\ is convex valued we fix $%
(x_{1},x_{2}),(x_{1},x_{2}^{\prime })\in \Gamma (\gamma )$ and $z$ s. t. $%
x_{2}<z<x_{2}^{\prime }$, and check that $\Gamma (\gamma )$ contains $%
(x_{1},z)$ too. Pick some $w\in \gamma (z)$: if $w>x_{1}$ we see that $%
\Gamma (\gamma )$ contains $(x_{1},x_{2})$ and $(w,z)$ s.t. $%
(x_{1},x_{2})\ll (w,z)$ which is a contradiction by Lemma 7.1. If $w<x_{1}$
we use instead $(w,z)$ and $(x_{1},x_{2}^{\prime })$ to reach a similar
contradiction, and we conclude $w=x_{1}$.

The proof below that $\gamma $ is single-valued a. e. will complete that~of
statement $ii)$.\smallskip

\textit{Statement} $iii)$ If $x_{1}<x_{1}^{\prime }$ in $\mathcal{X}$ and $%
\gamma ^{-}(x_{1})<\gamma ^{+}(x_{1}^{\prime })$ we again contradict the
strict supermodularity of $\mathcal{W}$ (Lemma 7.1) . So $%
x_{1}<x_{1}^{\prime }\Longrightarrow \gamma ^{-}(x_{1})\geq \gamma
^{+}(x_{1}^{\prime })$ and $\gamma ^{-}$\textit{\ }and $\gamma ^{+}$\ are
weakly decreasing\textit{.}

If $\gamma (x_{1})$ is not a singleton, $\gamma ^{+}(x_{1})>\gamma
^{-}(x_{1})$, then $\gamma ^{+}$ jumps down at $x_{1}$; a weakly decreasing
function can only do this a countable number of times. That the u.h.c.
closure of\textit{\ }$\gamma ^{+}$ contains $[\gamma ^{-}(x_{1}),\gamma
^{+}(x_{1})]$ follows from $\gamma ^{-}(x_{1})\geq \gamma ^{+}(x_{1}+\delta
) $ for any $\delta >0$.\smallskip

\textit{Statement }$iv)$ If $\gamma (L)$ does not contain $H$ we pick some $%
x_{1}$ in $\gamma (H)$: by statement $i)$ $\gamma (x_{1})$ contains $H$
therefore $x_{1}>L$; we reach a contradiction again from Lemma 7.1 because $%
\Gamma (\gamma )$ contains $(L,\gamma ^{+}(L))$ and the strictly larger $%
(x_{1},H)$.\smallskip

\textit{Statement} $v)$ Kakutani's theorem implies that at least one fixed
point exists. If $\Gamma (\gamma )$ contains both $(a,a)$ and $(b,b)$ we
contradicts again Lemma 7.1. Check finally that the inequalities $\gamma
^{-}(a)<a<\gamma ^{+}(a)$ are not compatible. Pick $\delta >0$ s.t. $\gamma
(a)$ contains $a-\delta $ and $a+\delta $: then $\Gamma (\gamma )$ contains $%
(a,a+\delta )$ and $(a-\delta ,a)$ (by symmetry) and we invoke Lemma 7.1
again.

\subsection{Differentiability of tight guarantees}

\textbf{Lemma 9.1 }\textit{Inheritance of differentiability }

\textit{Suppose} $\mathcal{X}=[L,H]$ \textit{is the interval }$L\leq x\leq H$
\textit{in }$%
\mathbb{R}
^{A}$\textit{. We fix} $x_{i}\in \mathcal{X}$,\textit{\ a tight guarantee} $%
g\in \mathcal{G}^{\varepsilon }$ for $\varepsilon =+,-$ \textit{and a
contact profile }$x=(x_{i},x_{-i})$ \textit{of }$g$\textit{\ at }$x_{i}$.%
\textit{\ If} $g(\cdot )$ \textit{and }$\mathcal{W}(\cdot ,x_{-i})$ \textit{%
are both differentiable at }$x_{i}$\textit{, we have}

\noindent $i)$ if $L<x_{i}<H$ then $\frac{dg}{dx_{i}}(x_{i})=\frac{\partial 
\mathcal{W}}{\partial x_{i}}(x_{i},x_{-i})$.

\noindent $ii)$ if $x_{i}=L$ and $g\in \mathcal{G}^{-}$, or $x_{i}=H$ and $%
g\in \mathcal{G}^{+}$ then $\frac{dg}{dx_{i}}(x_{i})\leq \frac{\partial 
\mathcal{W}}{\partial x_{i}}(x_{i},x_{-i})$.

\noindent $iii)$ if $x_{i}=H$ and $g\in \mathcal{G}^{-}$, or $x_{i}=L$ and $%
g\in \mathcal{G}^{+}$ then $\frac{dg}{dx_{i}}(x_{i})\geq \frac{\partial 
\mathcal{W}}{\partial x_{i}}(x_{i},x_{-i})$.

\textbf{Proof} \textit{Statement i). }Pick an arbitrary contact profile $x$
of $\mathcal{W}$ and $g\in \mathcal{G}^{-}$. Lemma 3.3 implies $%
g(x_{i}^{\ast })-g(x_{i})\leq \mathcal{W}(x_{i}^{\ast },x_{-i})-\mathcal{W}%
(x_{i},x_{-i})$ for all $x_{i}^{\ast }$ in some neighborhood of $x_{i}$. If $%
L<x_{i}<H$ and both $g(\cdot )$ and $\mathcal{W}(\cdot ,x_{-i})$ are
differentiable at $x_{i}$ we develop this inequality as $(\frac{dg}{dx_{i}}%
(x_{i})+o(x_{i}^{\ast }-x_{i}))\times (x_{i}^{\ast }-x_{i})\leq (\frac{%
\partial \mathcal{W}}{\partial x_{i}}(x)+o^{\prime }(x_{i}^{\ast
}-x_{i}))\times (x_{i}^{\ast }-x_{i})$ where both $o(\cdot )$ and $o^{\prime
}(\cdot )$ are continuous and vanish at zero. As $x_{i}^{\ast }-x_{i}$ can
take both signs, this implies the desired equality. The two inequalities in
statements $ii)$ and $iii)$ follow similarly when the sign of $x_{i}^{\ast
}-x_{i}$ is constant in the neighborhood of $x_{i}$. $\blacksquare
\smallskip $

Note that if $\mathcal{W}(\cdot ,x_{-i})$ is differentiable at $x_{i}$ the
function $z_{i}\rightarrow \mathcal{W}(z_{i},x_{-i})$ is $K$-Lipschitz for
some $K$ in a neighborhood $V$ of $x_{i}$. By the Corollary to Lemma 3.3, so
is $g(\cdot )$ at $x_{i}$ in $V$. In turn this implies that $g$ has bounded
variation on any interval, hence is differentiable in $x_{i}$ almost
everywhere in $[L,H]$.\smallskip

\textbf{Corollary} \textit{Suppose }$\mathcal{W}$ \textit{is differentiable
in }$[L,H]^{[n]}$\textit{. Then for} $\varepsilon =+,-$ \textit{the} \textit{%
tight guarantees in }$\mathcal{G}^{\varepsilon }$\textit{\ are characterised
by their contact set} $\mathcal{C}(g)$\textit{: for any two different }$%
g,h\in \mathcal{G}^{\varepsilon }$\textit{\ we have} $\mathcal{C}(g)\neq 
\mathcal{C}(h)$.

\textit{Moreover any (true) convex combination of }$g,h$ \textit{stays in }$%
\boldsymbol{G}^{\varepsilon }$\textit{\ but leaves }$\mathcal{G}%
^{\varepsilon }$: $]g,h[\cap \mathcal{G}^{\varepsilon }=\varnothing $\textit{%
.\smallskip }

\textbf{Proof}. By Lemma 9.1 if $\mathcal{C}(g)=\mathcal{C}(h)$ we get $%
\frac{dg}{dx}=\frac{dh}{dx}$ in the interval $]L,H[$ so $g$ and $h$ differ
by a constant, and if the constant is not zero one of $g,h$ is not tight.

For the second statement suppose that\textit{\ }$\mathcal{G}^{-}$ contains $%
g,h$ and $\frac{1}{2}(g+h)$, all different. Fix $x_{i}\in ]L,H[$ and a
contact profile $(x_{i},\widetilde{x}_{-i})$ of $\frac{1}{2}(g+h)$ at $x_{i}$%
. Clearly $\widetilde{x}_{-i}$ is also a contact profile of $g$ and of $h$
at $x_{i}$. Again by Lemma 9.2 this implies $\frac{dg}{dx_{i}}(x_{i})=\frac{%
dh}{dx_{i}}(x_{i})=\partial _{i}\mathcal{W}(x_{i},\widetilde{x}_{-1})$
almost surely in $]L,H[$. We conclude that $g-h$ is a constant and get a
contradiction of $g\neq h$. The argument for larger convex combinations with
general weights is entirely similar. $\blacksquare $

\subsection{Theorem 7.1}

\textit{Step 0: The integral in (\ref{12}) is well defined}.

For a correspondence $\gamma $\ as in Lemma 7.2 the integral $%
\int_{a}^{x_{1}}\partial _{1}\mathcal{W}(t,\gamma (t))dt$ is the value of $%
\int_{a}^{x_{1}}\partial _{1}\mathcal{W}(t,f(t))dt$ for any single-valued
selection $f$ of $\gamma $: this is independent of the choice of $f$ because 
$\gamma $ is multi-valued only at a countable number of points and every
single-valued selection of $\gamma (x_{1})$ is a measurable
function.\smallskip

\textit{Statement }$ii)$ Fix $g\in \mathcal{G}^{-}$ and its contact
correspondence $\gamma $. By Lemma 9.2 $g$ is differentiable a. e. in $[L,H]$%
. Its derivative $\frac{dg}{dx}$ is $\frac{dg}{dx}(x_{1})=\partial _{1}%
\mathcal{W}(x_{1},x_{2})$ for any $x_{2}\in \gamma (x_{1})$. Therefore we
can write the RH of (\ref{12}) as $\partial _{1}\mathcal{W}(x_{1},\gamma
(x_{1}))$ without specifying a particular selection of $\gamma (x_{1})$.

Note that $g(a)=una(a)$ because $(a,a)\in \Gamma (\gamma )$. Now integrating
the differential equation above with this initial condition at $a$ gives the
desired representation (\ref{12}).\smallskip

\textit{Statement }$i)$

\noindent \textit{Step 1} Lemma 7.2 implies that $\Gamma (\gamma )$ is a
one-dimensional line connecting $(L,H)$ and $(H,L)$ that we can parametrise
by a smooth mapping $s\rightarrow (\xi _{1}(s),\xi _{2}(s))$ from $[0,1]$
into $[L,H]^{2}$ s.t. $\xi _{1}(\cdot )$ increases weakly from $L$ to $H$
and $\xi _{2}(\cdot )$ decreases weakly from $H$ to $L$. We can also choose
this mapping so that $\xi _{1}(\frac{1}{2})=\xi _{2}(\frac{1}{2})=a$, the
fixed point of $\gamma $.\footnote{%
If $a$\ is $0$ or $1$\ we check that (\ref{12}) defines the two canonical
incremental guarantees in Theorem 4.1.}

We fix an arbitrary selection $\gamma ^{\ast }$ of $\gamma $, an arbitrary $%
\overline{x}_{1}$ in $[L,H]$, and check the identity%
\begin{equation}
\int_{a}^{\overline{x}_{1}}\partial _{1}\mathcal{W}(t,\gamma
(t))dt+\int_{a}^{\gamma ^{\ast }(\overline{x}_{1})}\partial _{1}\mathcal{W}%
(t,\gamma (t))dt=\mathcal{W}(\overline{x}_{1},\gamma ^{\ast }(\overline{x}%
_{1}))-\mathcal{W}(a,a)  \label{14}
\end{equation}

We change the variable $t$ to $s$ by $t=\xi _{1}(s)$ in the former and by $%
t=\xi _{2}(s)$ in the latter. Next $\overline{s}$ is the parameter at which $%
{\large (}\xi _{1}(\overline{s}),\xi _{2}(\overline{s}){\large )=(}\overline{%
x}_{1},\gamma ^{\ast }(\overline{x}_{1}))$ and we rewrite the LH of (\ref{14}%
) as $\int_{\frac{1}{2}}^{\overline{s}}\partial _{1}\mathcal{W}(\xi
_{1}(s),\xi _{2}(s))\frac{\partial \xi _{1}}{\partial s}(s)ds+\int_{\frac{1}{%
2}}^{\overline{s}}\partial _{1}\mathcal{W}(\xi _{2}(s),\xi _{1}(s))\frac{%
\partial \xi _{2}}{\partial s}(s)ds$, where in each term $\partial _{1}%
\mathcal{W}(t,\gamma (t))$ we can pick a proper selection of the (possible)
interval because $(\xi _{1}(s),\xi _{2}(s))\in \Gamma (\gamma )$. As $%
\mathcal{W}(x_{1},x_{2})$ is symmetric in $x_{1},x_{2}$, we can replace the
second integral by $\int_{\frac{1}{2}}^{\overline{s}}\partial _{2}\mathcal{W}%
(\xi _{1}(s),\xi _{2}(s))\frac{\partial \xi _{2}}{\partial s}(s)ds$ and
conclude that the sum is precisely $\mathcal{W}(\xi _{1}(\overline{s}),\xi
_{2}(\overline{s}))-\mathcal{W}(\xi _{1}(\frac{1}{2}),\xi _{2}(\frac{1}{2}))=%
\mathcal{W}(\overline{x}_{1},\gamma ^{\ast }(\overline{x}_{1}))-\mathcal{W}%
(a,a)$.\smallskip

\noindent \textit{Step 2} Equation (\ref{12}) defines a lower guarantee $g$: 
$g(x_{1})+g(x_{2})\leq \mathcal{W}(x_{1},x_{2})$ for $x_{1},x_{2}\in \lbrack
L,H]$.

The identity (\ref{14}) amounts to $g(x_{1})+g(\gamma ^{\ast }(x_{1}))=%
\mathcal{W}(x_{1},\gamma ^{\ast }(x_{1}))$ for all $x_{1}$. If we prove that 
$g\in \boldsymbol{G}^{-}$ this will imply it is tight. Compute\newline
$g(x_{1})+g(x_{2})=\mathcal{W}(x_{1},\gamma ^{\ast
}(x_{1}))+g(x_{2})-g(\gamma ^{\ast }(x_{1}))=\mathcal{W}(x_{1},\gamma ^{\ast
}(x_{1}))+\int_{\gamma ^{\ast }(x_{1})}^{x_{2}}\partial _{1}\mathcal{W}%
(t,\gamma (t))dt$.

We are left to show $\int_{\gamma ^{\ast }(x_{1})}^{x_{2}}\partial _{1}%
\mathcal{W}(t,\gamma (t))dt\leq \mathcal{W}(x_{1},x_{2})-\mathcal{W}%
(x_{1},\gamma ^{\ast }(x_{1}))$. Assume without loss $x_{1}\leq x_{2}$ and
distinguish several cases by the relative positions of\ $a$ and $x_{1},x_{2}$
.\smallskip

Case 1: $a\leq x_{1}\leq x_{2}$, so that $\gamma ^{\ast }(x_{1})\leq a$. For
every $t\geq \gamma ^{\ast }(x_{1})$ property $iii)$ in Lemma 7.2 implies $%
\gamma ^{+}(t)\leq \gamma ^{-}(\gamma ^{\ast }(x_{1}))$ and $\gamma (\gamma
^{\ast }(x_{1}))$ contains $x_{1}$: therefore submodularity of $\mathcal{W}$
implies $\partial _{1}\mathcal{W}(t,\gamma (t))\leq \partial _{1}\mathcal{W}%
(t,x_{1})$ and\newline
$\int_{\gamma ^{\ast }(x_{1})}^{x_{2}}\partial _{1}\mathcal{W}(t,\gamma
(t))dt\leq \int_{\gamma ^{\ast }(x_{1})}^{x_{2}}\partial _{1}\mathcal{W}%
(t,x_{1})dt=\mathcal{W}(x_{2},x_{1})-\mathcal{W}(\gamma ^{\ast
}(x_{1}),x_{1})$.\smallskip

Case 2: $x_{1}\leq a\leq \gamma ^{\ast }(x_{1})\leq x_{2}$. Similarly for $%
t\geq \gamma ^{\ast }(x_{1})$ we have $\gamma ^{+}(t)\leq \gamma ^{-}(\gamma
^{\ast }(x_{1}))$ and conclude as in Case 1.\smallskip

Case 3: $x_{1}\leq x_{2}\leq a$, so that $\gamma ^{\ast }(x_{1})\geq a$. For
all $t\leq \gamma ^{\ast }(x_{1})$ we have $\gamma ^{-}(t)\geq \gamma
^{+}(\gamma ^{\ast }(x_{1}))$ and $\gamma (\gamma ^{\ast }(x_{1}))$ contains 
$x_{1}$: now submodularity of $\mathcal{W}$ gives $\partial _{1}\mathcal{W}%
(t,z)\geq \partial _{1}\mathcal{W}(t,x_{2})$ for $z$ between $x_{2}$ and $%
\gamma ^{\ast }(x_{1})$ and the desired inequality because the integral $%
\int_{\gamma ^{\ast }(x_{1})}^{x_{2}}$ goes from high to low.

Case 4: $x_{1}\leq a\leq x_{2}\leq \gamma ^{\ast }(x_{1})$. Same argument as
in Case 3.

\pagebreak

\includepdf{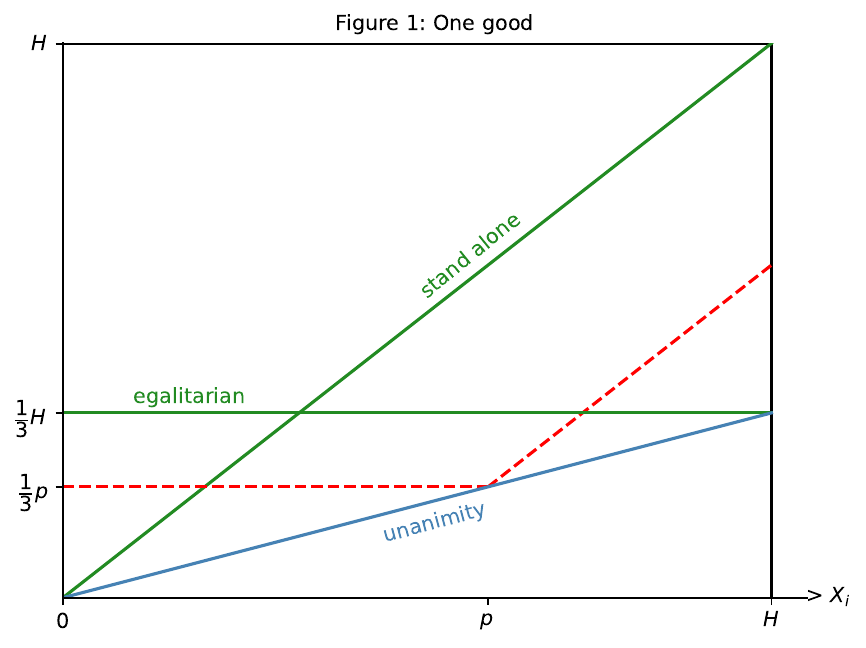}   
\pagebreak
\includepdf{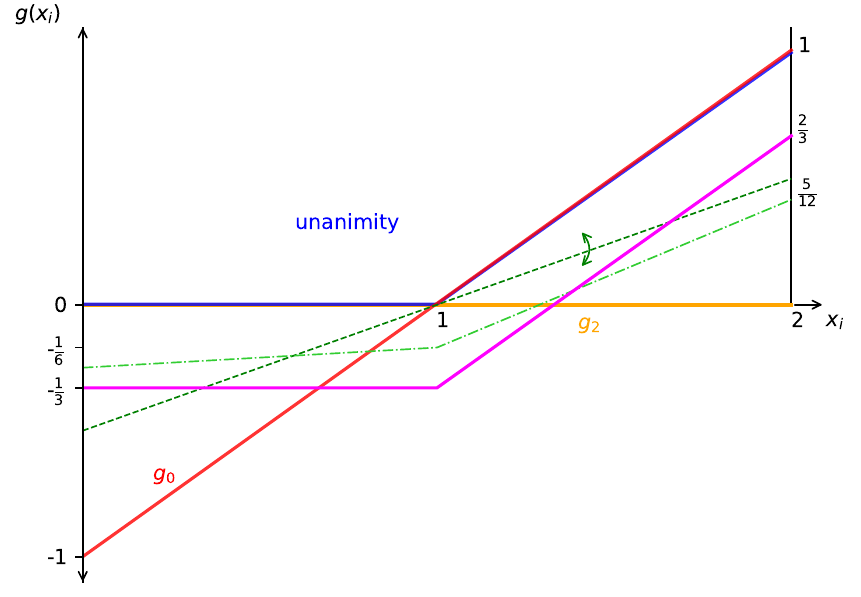}   

\end{document}